\documentclass[journal]{IEEEtran}
\usepackage[T1]{fontenc}
\usepackage[latin9]{inputenc}
\usepackage{float}
\usepackage{amsmath}
\usepackage{amssymb}
\usepackage{graphicx}
\usepackage[bookmarks=true,bookmarksnumbered=true,bookmarksopen=true,bookmarksopenlevel=1,
 breaklinks=false,pdfborder={0 0 0},pdfborderstyle={},backref=false,colorlinks=false]
 {hyperref}
\hypersetup{pdftitle={Your Title},
 pdfauthor={Your Name},
 pdfpagelayout=OneColumn, pdfnewwindow=true, pdfstartview=XYZ, plainpages=false}

\makeatletter

\newcommand{\lyxmathsym}[1]{\ifmmode\begingroup\def\b@ld{bold}
  \text{\ifx\math@version\b@ld\bfseries\fi#1}\endgroup\else#1\fi}

\providecommand{\tabularnewline}{\\}
\newcommand{\lyxdot}{.}

\floatstyle{ruled}
\newfloat{algorithm}{tbp}{loa}
\providecommand{\algorithmname}{Algorithm}
\floatname{algorithm}{\protect\algorithmname}

\usepackage[caption=false,font=footnotesize]{subfig}

\usepackage{algorithm} 
\usepackage{algorithmic}

\makeatother

\begin{document}
\title{Antenna Coding Empowered by Pixel Antennas}
\author{Shanpu~Shen,~\IEEEmembership{Senior Member,~IEEE}, Kai-Kit Wong,~\IEEEmembership{Fellow,~IEEE},
and Ross Murch,~\IEEEmembership{Fellow,~IEEE}\thanks{Manuscript received; This work was funded by the Science and Technology
Development Fund, Macau SAR (File/Project no. 001/2024/SKL) and the
Hong Kong Research Grants Council Area of Excellence Grant AoE/E-601/22-R.
\textit{(Corresponding author: Shanpu Shen.)}}\thanks{S. Shen is with the State Key Laboratory of Internet of Things for
Smart City and Department of Electrical and Computer Engineering,
University of Macau, Macau, China (email: shanpushen@um.edu.mo).}\thanks{K. K. Wong is affiliated with the Department of Electronic and Electrical
Engineering, University College London, Torrington Place, WC1E 7JE,
United Kingdom and he is also affiliated with Yonsei Frontier Lab,
Yonsei University, Seoul, Korea (e-mail: kai-kit.wong@ucl.ac.uk).}\thanks{R. Murch is with the Department of Electronic and Computer Engineering
and the Institute of Advanced Study, The Hong Kong University of Science
and Technology, Clear Water Bay, Kowloon, Hong Kong (e-mail: eermurch@ust.hk).}}
\maketitle
\begin{abstract}
Pixel antennas, based on discretizing a continuous radiation surface
into small elements called pixels, are a flexible reconfigurable antenna
technology. By controlling the connections between pixels via switches,
the characteristics of pixel antennas can be adjusted to enhance the
wireless channel. Inspired by this, we propose a novel technique denoted
antenna coding empowered by pixel antennas. We first derive a physical
and electromagnetic based communication model for pixel antennas using
microwave multiport network theory and beamspace channel representation.
With the model, we optimize the antenna coding to maximize the channel
gain in a single-input single-output (SISO) pixel antenna system and
develop a codebook design for antenna coding to reduce the computational
complexity. We analyze the average channel gain of SISO pixel antenna
system and derive the corresponding upper bound. In addition, we jointly
optimize the antenna coding and transmit signal covariance matrix
to maximize the channel capacity in a multiple-input multiple-output
(MIMO) pixel antenna system. Simulation results show that using pixel
antennas can enhance the average channel gain by up to 5.4 times and
channel capacity by up to 3.1 times, demonstrating the significant
potential of pixel antennas as a new dimension to design and optimize
wireless communication systems.
\end{abstract}


\section{Introduction}

\IEEEPARstart{A}{ntennas} play a significant role in wireless communication
systems. Utilizing multiple antennas, multiple-input multiple-output
(MIMO) systems have been one of the most celebrated wireless communication
technologies over the past few decades. It has evolved from point-to-point
MIMO in the third generation (3G) to multiuser MIMO in the fourth
generation (4G) and massive MIMO in the fifth generation (5G), and
will still be a dominating technology in the upcoming sixth generation
(6G) \cite{10379539}. However, the antennas utilized in conventional
MIMO technologies have fixed configurations and characteristics such
as operating frequency, radiation pattern, and polarization, which
are not involved in the signal processing and system optimization
in wireless communications. As 6G mobile communication strives to
push the performance to extreme levels far beyond current 5G mobile
communications, it is necessary to seek a paradigm shift in antenna
technology to achieve a new degree of freedom in designing and optimizing
wireless communication systems, so as to break through the current
performance limit.

Pixel antennas are a flexible antenna technology to design highly
reconfigurable antennas and dates back to 2004 \cite{1367557}, \cite{1303859}.
The concept of the pixel antenna is based on discretizing a continuous
radiation surface into small elements, which are referred to as pixels,
and connecting adjacent pixels through hardwires or RF switches \cite{5991914,6163369,6781015},
\cite{Pixel_Song}, \cite{7950976}, \cite{9743796} and \cite{9769906}.
By controlling the connections between pixels, various antenna topologies
can be formed so that the pixel antenna can be reconfigured to achieve
a wide range of distinct antenna characteristics such as operating
frequency, radiation pattern, and polarization. For example, a frequency-reconfigurable
pixel antenna, which can reconfigure the operating frequency from
PCS band to UMTS band, has been designed in \cite{Pixel_Song}, and
pattern-reconfigurable pixel antennas, which can reconfigure the radiation
pattern to implement $360\lyxmathsym{\textdegree}$ single- or multi-beam
steering, have been designed in \cite{7950976}, \cite{9743796},
and \cite{9769906}. Moreover, various pixel antennas have been designed
for different applications. In \cite{9852001}, a reconfigurable intelligent
surface with pixelated elements, which can reconfigure the phase shift
of reflected waves, has been designed. In \cite{ShanpuShen2017_TAP_EHPIXEL},
and \cite{9233414}, multiport pixel rectennas, which can maximize
the output dc power for ambient RF energy harvesting, have been developed.
In \cite{9506812}, MIMO antennas with pixelated surface, which provide
low mutual coupling and spatial correlation to maximize the ergodic
capacity, have been developed. In \cite{9096388}, defected ground
structures with pixelated grid, which suppress cross-polarization
and achieve circular polarization, have been designed. While these
works \cite{1303859}, \cite{5991914,6163369,6781015}, \cite{Pixel_Song},
\cite{7950976}, \cite{9743796}, \cite{9769906}, \cite{9852001},
\cite{ShanpuShen2017_TAP_EHPIXEL}, \cite{9233414}, \cite{9506812},
\cite{9096388} have designed pixel antennas with significant performance
enhancement, the investigations are limited at the level of antenna
hardware design and do not consider the impact on wireless communications
at the level of the system. Thus, it is necessary to explore using
pixel antennas to design and optimize wireless communication systems.

In parallel with the development of pixel antennas, the concept of
the fluid antenna system (FAS) has also been proposed and introduced
in 2020 \cite{9131873}. FAS originated from the emergence of mechanically
flexible antennas such as liquid antennas based on liquid metals or
ionized solutions \cite{9388928}, which enables a single antenna
freely switching the position in a small linear space. Moreover, FAS
can be implemented by any appropriate reconfigurable technique including
electronically flexible pixel antennas, where the pixels can be turned
on-and-off instantly so that several pixels can be turned on to form
an antenna port in a particular position. One of the first pixel antenna
designs for an FAS has been demonstrated in \cite{zhang2024pixel}
to show that a single pixel antenna can mimic switching positions
in a small linear space and be useful for FAS. Leveraging the position-switchable
property, FAS can select the optimal position in a given space to
adapt to fading channels \cite{9264694} and thus enhance the system
performance such as level crossing rate \cite{9131873}, diversity
gain \cite{10188603}, and outage probability \cite{10130117}. FAS
has also been utilized to assist MIMO communication systems \cite{10303274},
\cite{10243545}, \cite{zheng2023flexibleposition}, \cite{10328751}.
Specifically, the capacity, diversity-multiplexing tradeoff, spectral
efficiency, and energy efficiency for MIMO-FAS have been investigated
in \cite{10303274}, \cite{10243545}, \cite{zheng2023flexibleposition}
and the rate maximization for MIMO-FAS based on statistical channel
state information (CSI) has been studied in \cite{10328751}. In addition,
a novel multiple access based on FAS, denoted as fluid antenna multiple
access (FAMA), has been proposed in \cite{9650760}, and FAMA can
be categorized into fast FAMA \cite{9953084} and slow FAMA \cite{10066316},
Moreover, FAS has been investigated to enhance emerging areas such
as secret communications \cite{10092780} and full duplex communications
\cite{10184308}.

While the emergence and development of FAS \cite{9131873}, \cite{9264694},
\cite{10188603}, \cite{10130117}, \cite{10303274}, \cite{10243545},
\cite{zheng2023flexibleposition}, \cite{10328751}, \cite{9650760},
\cite{9953084}, \cite{10066316}, \cite{10092780}, \cite{10184308}
has preliminarily demonstrated the potential of pixel antennas for
designing and optimizing wireless communication system, there is much
further potential for pixel antennas to be exploited in wireless communication.
For example, utilizing the individual pixels inside a pixel antenna,
and the connections between them, can open up new possibilities for
their use in wireless communication. To fully exploit the potential
of pixel antennas for wireless communication, there remains two open
problems: 1) how to derive a physical and electromagnetic (EM) based
communication model for pixel antennas, and 2) how to efficiently
utilize the flexibility of pixel antennas to enhance wireless communications.

In this paper, we propose using pixel antennas to enhance wireless
communication systems with an EM based model and a systematic and
efficient approach. In comparison with conventional wireless communication
systems using antennas with fixed configuration, using pixel antennas
allows designing and optimizing wireless communication systems in
a new dimension through adjusting the antenna configuration, and thus
the wireless communication systems performance can be significantly
enhanced. The contributions of the paper are summarized as follows.

\textit{First}, we propose a novel technique called \textit{antenna
coding} empowered by pixel antennas to enhance wireless communication
systems. By adjusting the connections between adjacent pixels through
switches, the characteristics of the pixel antenna such as impedance,
radiation pattern, and polarization can be controlled. Inspired by
this, we propose antenna coding, which is a technique to control antenna
configuration and characteristics through binary codes representing
switches for enhancing wireless systems. With antenna coding, wireless
systems can be optimized with a new dimension and thus achieve enhanced
performance compared to conventional systems with fixed antenna configuration.

\textit{Second}, we derive a physical and EM based model for pixel
antennas. Using microwave multiport network theory, we analyze the
pixel antenna and model its configuration and characteristics such
as radiation pattern as functions of a binary vector referred to as
the \textit{antenna coder} which represents the switch on and off
states. Further, using the beamspace channel representation, we incorporate
radiation patterns into the channel to derive a communication model
for pixel antennas, where the channel can be controlled by antenna
coder, allowing antenna coding optimization to enhance wireless systems.

\textit{Third}, we optimize the antenna coding to maximize the channel
gain in single-input single-output (SISO) pixel antenna systems. The
antenna coding optimization is a binary optimization problem and can
be solved by a successive exhaustive Boolean optimization approach.
To reduce the computational complexity for optimization, we propose
a codebook design for antenna coding and optimize the antenna coder
by searching the codebook.

\textit{Fourth}, we analyze the average channel gain of SISO pixel
antenna systems and find that the upper bound of average channel gain
is the effective aerial degrees-of-freedom of pixel antennas which
can be found by singular value decomposition of the open-circuit radiation
pattern matrix.

\textit{Fifth}, we jointly optimize the antenna coding and transmit
signal covariance matrix to maximize the channel capacity in MIMO
pixel antenna systems by using exhaustive Boolean optimization approach
and searching of the codebook design.

\textit{Sixth}, we evaluate the performance of SISO and MIMO pixel
antenna systems compared to conventional SISO and MIMO systems using
antennas with fixed configuration. The simulation results show that
using pixel antennas can enhance the average channel gain by up to
5.4 times and the channel capacity by up to 3.1 times, demonstrating
the significant potential of pixel antennas in wireless communication
systems.

\textit{Organization}: Section II introduces the pixel antenna model
and antenna coding. Section III presents the SISO pixel antenna system
and antenna coding design for channel gain maximization. Section IV
presents the MIMO pixel antenna system and antenna coding design for
capacity maximization. Section V provides performance evaluation for
the SISO and MIMO pixel antenna system. Section VI provides discussions
and future work. Section VII concludes this work.

\textit{Notation}: Bold lower and upper case letters represent vectors
and matrices, respectively. A symbol without bold font denotes a scalar.
$\mathbb{R}$ and $\mathbb{C}$ represent real and complex number
sets, respectively. $\mathbb{E}\left[\cdotp\right]$ denotes the expectation.
$\left(\cdotp\right)^{+}$denotes the positive part. $\left|a\right|$
denotes the modulus of a complex number $a$. $\left[\mathbf{a}\right]_{i}$
and $\left\Vert \mathbf{a}\right\Vert $ denote the $i$th element
and $l_{2}$-norm of a vector $\mathbf{a}$, respectively. $\mathbf{A}^{*}$,
$\mathbf{A}^{T}$, $\mathbf{A}^{H}$, $\left[\mathbf{A}\right]_{:,i}$,
$\left[\mathbf{A}\right]_{i,j}$, $\left|\mathbf{A}\right|$, and
$\textrm{Tr}\left(\mathbf{A}\right)$ denote the conjugate, transpose,
conjugate transpose, $i$th column, $\left(i,j\right)$th element,
determinant, and trace of a matrix $\mathbf{A}$, respectively. $\mathcal{CN}(\boldsymbol{0},\boldsymbol{\Sigma})$
denotes the circularly symmetric complex Gaussian distribution with
mean \textbf{0} and covariance matrix $\boldsymbol{\Sigma}$. $\text{diag}\left(a_{1},...,a_{N}\right)$
denotes a diagonal matrix with entries $a_{1},...,a_{N}$. $\mathbf{I}$
denotes an identity matrix.

\section{Pixel Antenna}

\begin{figure}[t]
\begin{centering}
\includegraphics[width=8.8cm]{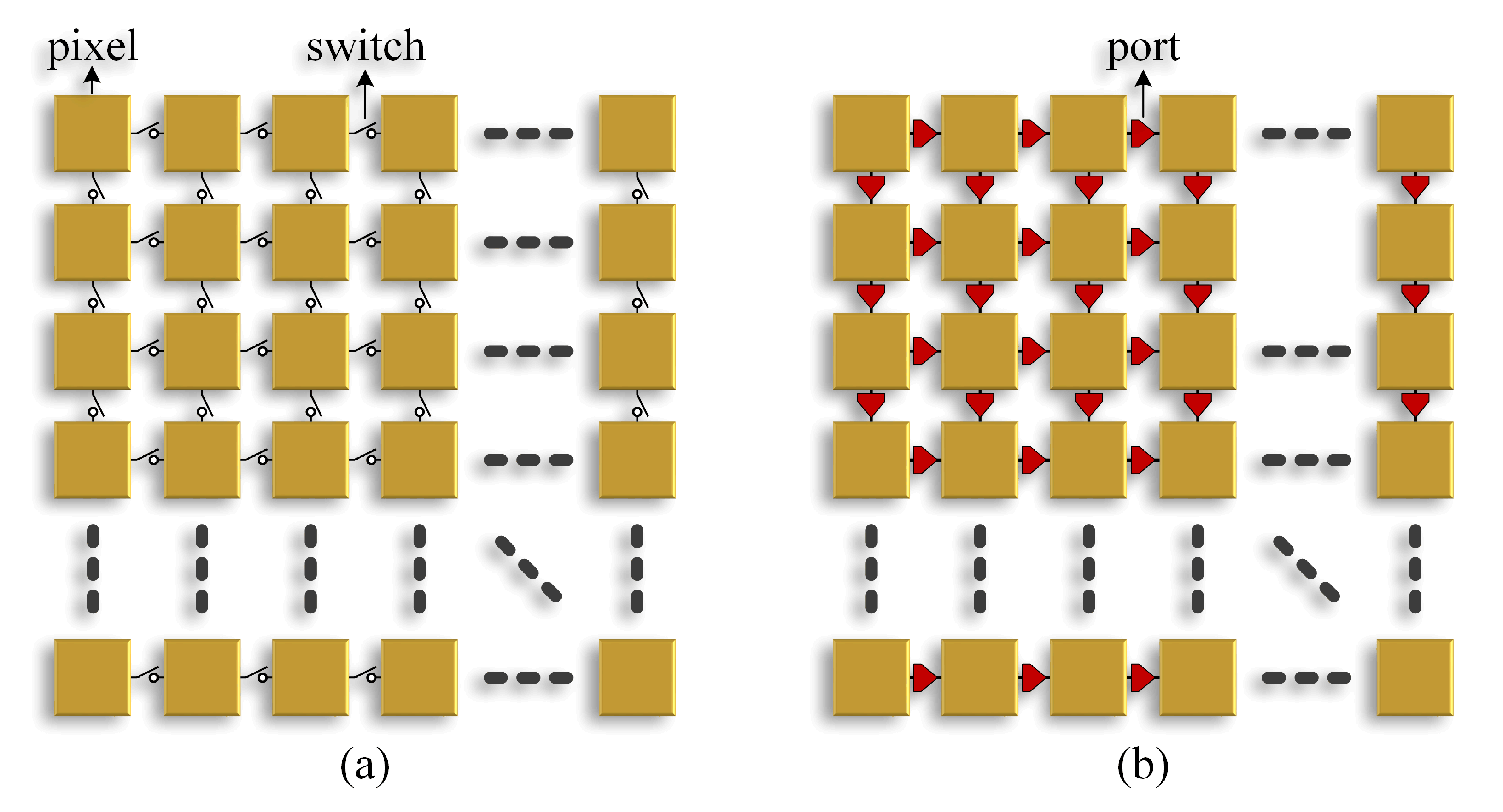}
\par\end{centering}
\caption{\label{fig: pixel antenna schematics}(a) Schematic for pixel antenna.
(b) Multiport circuit network for pixel antenna. Red arrows represent
the ports which replace switches.}
\end{figure}

\begin{figure}[t]
\begin{centering}
\includegraphics[width=8cm]{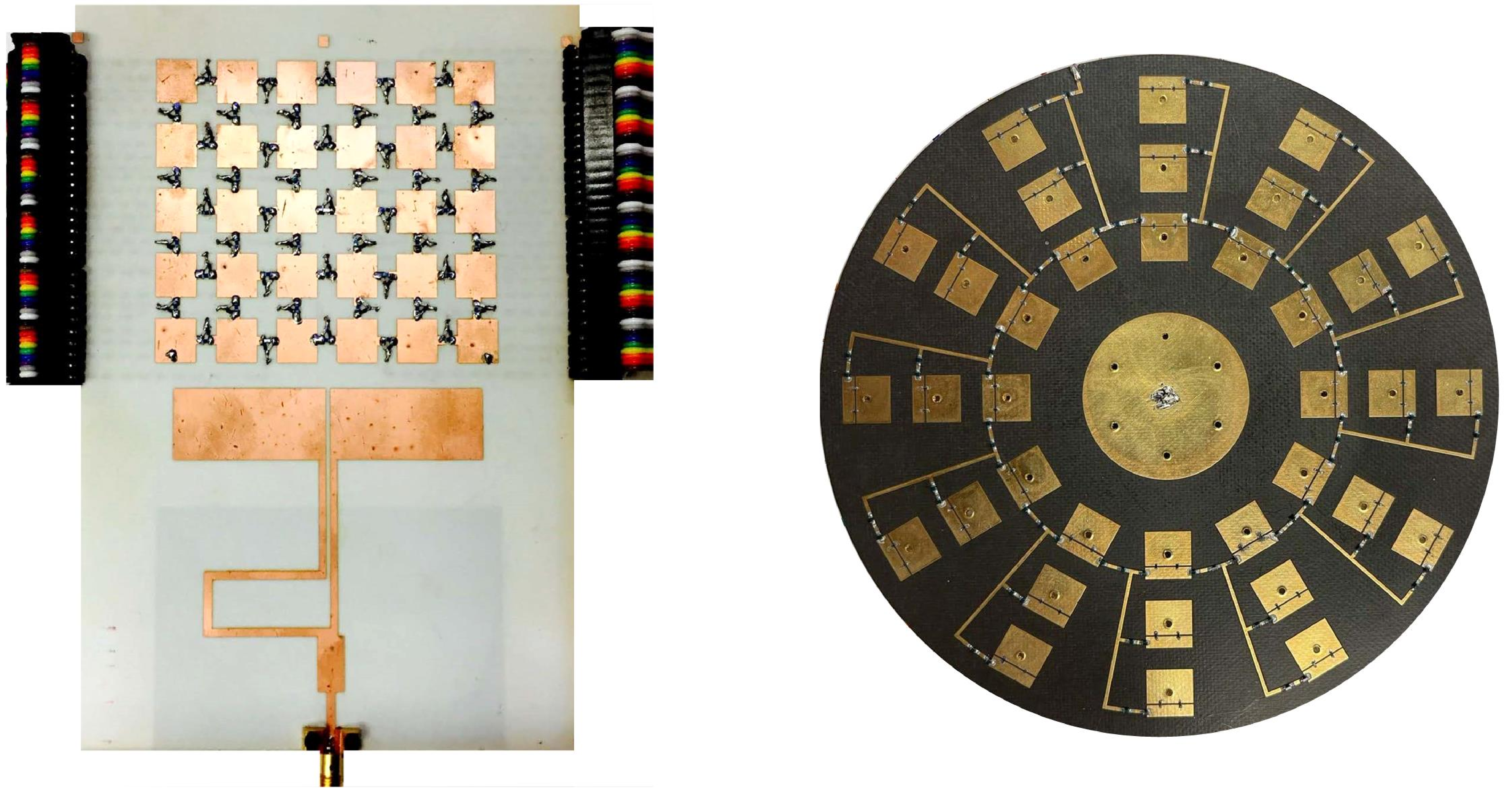}
\par\end{centering}
\caption{\label{fig:pixel antenna example}Two examples of pixel antenna designs
proposed in \cite{7950976} and \cite{9769906}.}
\end{figure}

\begin{figure}[t]
\begin{centering}
\includegraphics[width=7.5cm]{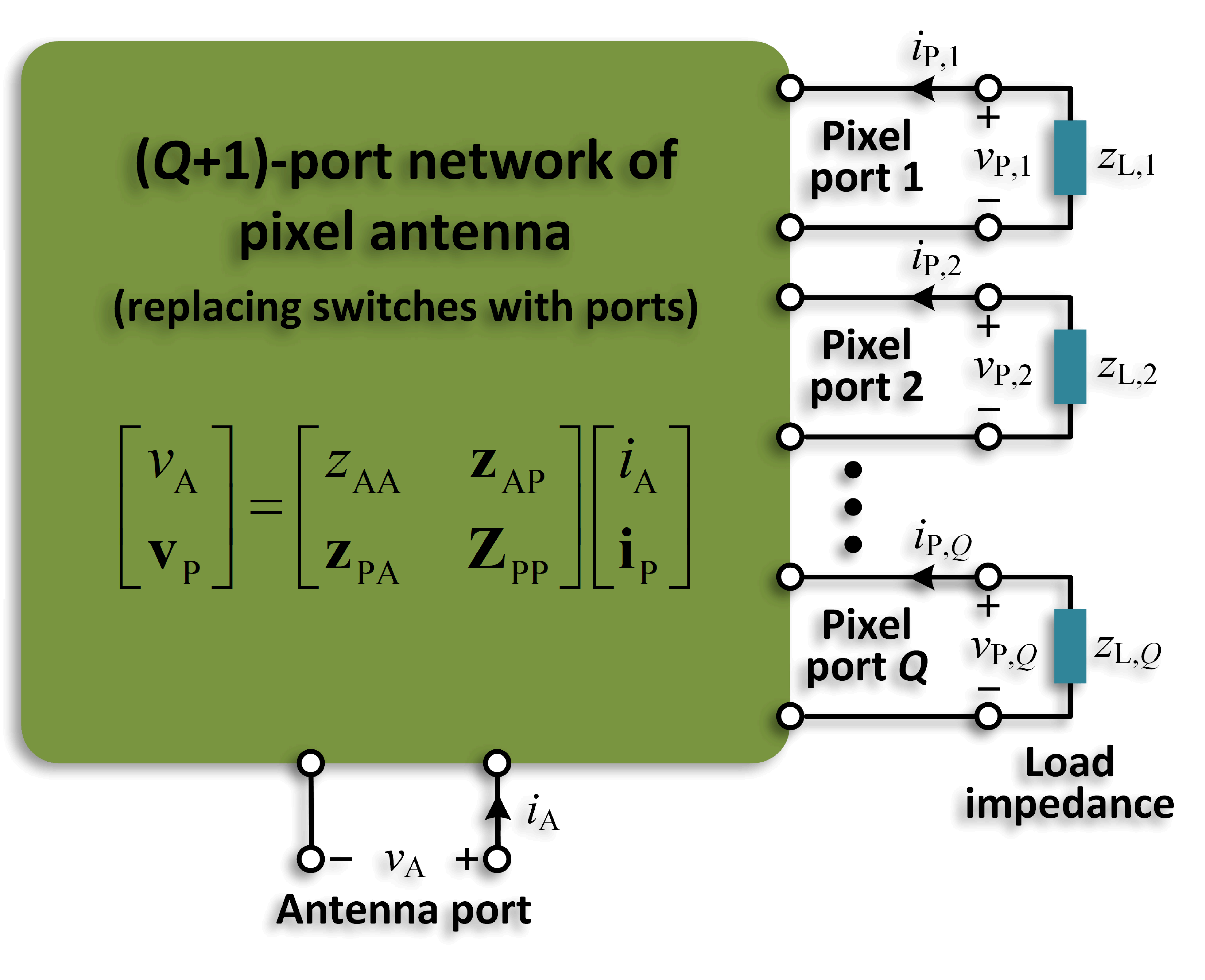}
\par\end{centering}
\caption{\label{fig:pixel antenna multiport}Model of the multiport circuit
network for pixel antenna.}
\end{figure}

\begin{figure*}[tbh]
\begin{centering}
\includegraphics[width=17.5cm]{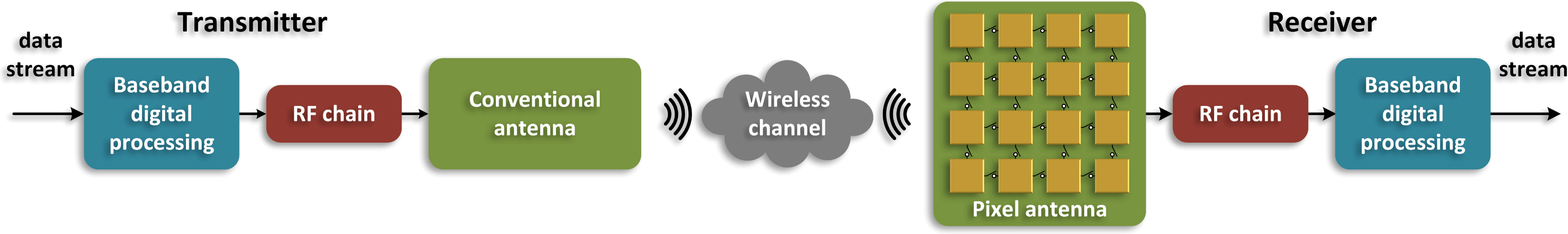}
\par\end{centering}
\caption{\label{fig:SISO PAS schematic}System diagram of the SISO pixel antenna
system.}
\end{figure*}

In this section, we introduce the pixel antenna model and the proposed
antenna coding technique. As illustrated in the schematic in Fig.
\ref{fig: pixel antenna schematics}(a), the pixel antenna is based
on discretizing a continuous radiation surface into small elements,
which are referred to as pixels, and adjacent pixels can be connected
through switches to flexibly control the antenna configuration and
characteristics such as radiation pattern. Two examples of pattern-reconfigurable
pixel antenna designs proposed in \cite{7950976} and \cite{9769906}
are shown in Fig. \ref{fig:pixel antenna example}.

The pixel antenna can be modeled using microwave multiport network
theory\footnote{The multiport circuit network model can be traced back to the method
of moments (MoM) \cite{Johnson1999} which uses an impedance matrix
to model an antenna. Using multiport circuit network to model lumped
elements, such as capacitor, inductor, resistor, and RF switches,
has been developed and demonstrated in \cite{PerruisseauCarrier2010},
\cite{Yousefbeiki2012} and verified in reconfigurable antenna designs
\cite{Pixel_Song}, \cite{LeonardoAraqueQuijano2009}.}. Specifically, considering a pixel antenna which has $Q$ switches,
we can model the pixel antenna as a multiport circuit network by replacing
the $Q$ switches with $Q$ ports as shown in Fig. \ref{fig: pixel antenna schematics}(b).
Thus, the entire pixel antenna can be modeled as a $(Q+1)$-port circuit
network, which consists of one antenna port and $Q$ pixel ports denoting
the connections between adjacent pixels (which are numbered 2 to $Q+1$),
as shown in Fig. \ref{fig:pixel antenna multiport}. The $(Q+1)$-port
circuit network can be characterized by a $(Q+1)\times(Q+1)$ impedance
matrix $\mathbf{Z}$, which can be partitioned as
\begin{equation}
\mathbf{Z}=\left[\begin{array}{ll}
z_{\textrm{AA}} & \mathbf{z}_{\textrm{AP}}\\
\mathbf{z}_{\textrm{PA}} & \mathbf{Z}_{\textrm{PP}}
\end{array}\right],\label{eq:Z-one_sim}
\end{equation}
where $z_{\textrm{AA}}\in\mathbb{C}$ is the self impedance for the
antenna port, $\mathbf{Z}_{\textrm{PP}}\in\mathbb{C}^{Q\times Q}$
is the impedance matrix for the $Q$ pixel ports, $\mathbf{z}_{\textrm{AP}}\in\mathbb{C}^{1\times Q}$
is the trans-impedance relating the voltage of the antenna port with
currents of the $Q$ pixel ports, and $\mathbf{z}_{\textrm{PA}}\in\mathbb{C}^{Q\times1}$
is the transpose of $\mathbf{z}_{\textrm{AP}}$. We denote the voltages
at the antenna port and pixel ports as $v_{\textrm{A}}\in\mathbb{C}$
and $\mathbf{v}_{\textrm{P}}=\left[v_{\textrm{P},1},\ldots,v_{\textrm{P},Q}\right]\in\mathbb{C}^{Q\times1}$,
respectively, and the currents at the antenna port and pixel ports
as $i_{\textrm{A}}\in\mathbb{C}$ and $\mathbf{i}_{\textrm{P}}=\left[i_{\textrm{P},1},\ldots,i_{\textrm{P},Q}\right]\in\mathbb{C}^{Q\times1}$,
respectively. Using \eqref{eq:Z-one_sim}, we can relate the voltages
and currents by
\begin{equation}
\left[\begin{array}{c}
v_{\textrm{A}}\\
\mathbf{v}_{\textrm{P}}
\end{array}\right]=\left[\begin{array}{ll}
z_{\textrm{AA}} & \mathbf{z}_{\textrm{AP}}\\
\mathbf{z}_{\textrm{PA}} & \mathbf{Z}_{\textrm{PP}}
\end{array}\right]\left[\begin{array}{c}
i_{\textrm{A}}\\
\mathbf{i}_{\textrm{P}}
\end{array}\right].\label{eq:V=00003DZI}
\end{equation}

The switch connected to the $q$th pixel port has two states, on or
off\footnote{It is the RF switch that has on or off states and that it is not the
pixel element that has on or off states. The RF switch in off state
will only make the current at the RF switch zero, but this does not
mean that the surface current through the pixel element will disappear,
which is because pixel elements are closely coupled with each other.
In addition, control of the RF switch states can be achieved by ensuring
a dc path exists through the structure by using RF chokes.}. As shown in Fig. \ref{fig:pixel antenna multiport}, we can model
the switch as a load impedance $z_{\textrm{L},q}$ $\forall q$ with
open or short circuit, i.e. $z_{\textrm{L},q}=0$ or $\textrm{\ensuremath{\infty}}$.
Therefore, a binary variable $b_{q}\in\left\{ \textrm{0,1}\right\} $
can be used to denote the state of the switch connected to the $q$th
port, i.e.
\begin{equation}
z_{\textrm{L},q}=\begin{cases}
0, & b_{q}=0,\:\textrm{i.e. switch\:on,}\\
\infty, & b_{q}=1,\:\textrm{i.e. switch\:off}.
\end{cases}\label{eq:map b and zl}
\end{equation}
Numerically, for the switch off state, we can set $z_{\textrm{L},q}$
as a very large value, for example $z_{\textrm{L},q}=10^{8}$, to
approach the infinite. We group $z_{\textrm{L},q}$ $\forall q$ into
a diagonal load impedance matrix $\mathbf{Z}_{\textrm{L}}=\textrm{diag}(z_{\textrm{L},1},\ldots,z_{\textrm{L},Q})\in\mathbb{C}^{Q\times Q}$
and group $b_{q}$ $\forall q$ into a vector $\mathbf{b}=[b_{1},\ldots,b_{Q}]^{T}\in\mathbb{R}^{Q\times1}$,
which represents the states of all switches and is referred to as
an antenna coder. Therefore, the load impedance matrix $\mathbf{Z}_{\textrm{L}}$
can be coded by $\mathbf{b}$, denoted as $\mathbf{Z}_{\textrm{L}}\left(\mathbf{b}\right)$,
so that $\mathbf{v}_{\textrm{P}}$ and $\mathbf{i}_{\textrm{P}}$
can be related by
\begin{equation}
\mathbf{v}_{\textrm{P}}=-\mathbf{Z}_{\textrm{L}}\left(\mathbf{b}\right)\mathbf{i}_{\textrm{P}}.\label{eq:V=00003DZloadI}
\end{equation}
Substituting \eqref{eq:V=00003DZloadI} into \eqref{eq:V=00003DZI},
we can find that the currents at the pixel ports can be coded by the
antenna coder $\mathbf{b}$, written as
\begin{equation}
\mathbf{i}_{\textrm{P}}\left(\mathbf{b}\right)=-\left(\mathbf{Z}_{\textrm{PP}}+\mathbf{Z}_{\textrm{L}}\left(\mathbf{b}\right)\right)^{-1}\mathbf{z}_{\textrm{PA}}i_{\textrm{A}}.\label{eq:current_relation}
\end{equation}

Accordingly, the radiation pattern of the pixel antenna can also be
coded by the antenna coder $\mathbf{b}$. Given an antenna coder $\mathbf{b}$,
we denote the radiation pattern of the pixel antenna excited by current
$i_{\textrm{A}}$ as $\mathbf{e}\left(\mathbf{b}\right)=[\mathbf{e}_{\theta}^{T}\left(\mathbf{b}\right),\mathbf{e}_{\phi}^{T}\left(\mathbf{b}\right)]^{T}\in\mathbb{C}^{2K\times1}$
where $\mathbf{e}_{\theta}\left(\mathbf{b}\right),\mathbf{e}_{\phi}\left(\mathbf{b}\right)\in\mathbb{C}^{K\times1}$
are $\theta$ and $\phi$ polarization components over $K$ sampled
spatial angles, respectively. The radiation pattern of the pixel antenna
is the superposition of the radiation patterns of antenna and pixel
ports weighted by currents. Thus we can express $\mathbf{e}\left(\mathbf{b}\right)$
as
\begin{equation}
\mathbf{e}\left(\mathbf{b}\right)=\mathbf{e}_{\textrm{A}}i_{\textrm{A}}+\sum_{q=1}^{Q}\mathbf{e}_{\textrm{P},q}i_{\textrm{P},q}\left(\mathbf{b}\right)=\mathbf{E}_{\textrm{oc}}\mathbf{i}\left(\mathbf{b}\right),\label{eq:F pattern}
\end{equation}
where $\mathbf{e}_{\textrm{A}}=[\mathbf{e}_{\textrm{A},\theta}^{T},\mathbf{e}_{\textrm{A},\phi}^{T}]^{T}\in\mathbb{C}^{2K\times1}$
is the radiation pattern (with $\mathbf{e}_{\textrm{A},\theta},\mathbf{e}_{\textrm{A},\phi}\in\mathbb{C}^{K\times1}$
being $\theta$ and $\phi$ polarization components) of the antenna
port excited by a unit current when all the other ports are open-circuited,
$\mathbf{e}_{\textrm{P},q}=[\mathbf{e}_{\textrm{P},q,\theta}^{T},\mathbf{e}_{\textrm{P},q,\phi}^{T}]^{T}\in\mathbb{C}^{2K\times1}$
is the radiation pattern (with $\mathbf{e}_{\textrm{P},q,\theta},\mathbf{e}_{\textrm{P},q,\phi}\in\mathbb{C}^{K\times1}$
being $\theta$ and $\phi$ polarization components) of the $q$th
pixel port excited by unit current when all the other ports are open-circuit,
the matrix $\mathbf{E}_{\textrm{oc}}=[\mathbf{e}_{\textrm{A}},\mathbf{e}_{\textrm{P},1},\ldots,\mathbf{e}_{\textrm{P},Q}]\in\mathbb{C}^{2K\times(Q+1)}$
collects the open-circuit radiation patterns of all ports and is named
as open-circuit radiation pattern matrix, and $\mathbf{i}\left(\mathbf{b}\right)\in\mathbb{C}^{(Q+1)\times1}$
collects the currents at all port, expressed as
\begin{equation}
\mathbf{i}\left(\mathbf{b}\right)=\left[\begin{array}{c}
i_{\textrm{A}}\\
\mathbf{i}_{\textrm{P}}\left(\mathbf{b}\right)
\end{array}\right]=\left[\begin{array}{c}
1\\
-\left(\mathbf{Z}_{\textrm{PP}}+\mathbf{Z}_{\textrm{L}}\left(\mathbf{b}\right)\right)^{-1}\mathbf{z}_{\textrm{PA}}
\end{array}\right]i_{\textrm{A}}.\label{eq: current}
\end{equation}
The load impedance matrix $\mathbf{Z}_{\textrm{L}}\left(\mathbf{b}\right)$
can be coded by the antenna coder $\mathbf{b}$, which enables coding
of the currents $\mathbf{i}\left(\mathbf{b}\right)$ and radiation
pattern of the pixel antenna $\mathbf{e}\left(\mathbf{b}\right)$.
Thus, given the impedance matrix $\mathbf{Z}$ and open-circuit radiation
pattern matrix $\mathbf{E}_{\textrm{oc}}$, we can find the radiation
pattern of pixel antenna $\mathbf{e}\left(\mathbf{b}\right)$ configured
by any antenna coder by \eqref{eq:F pattern} and \eqref{eq: current}
with low computational complexity of $\mathcal{O}\left(Q^{3}+Q^{2}+Q+2K(Q+1)\right)$.
In Section V-D, we will show that this multiport circuit network model
has the same accuracy as the full-wave EM simulation but requires
significantly less computational time.

In total, there are $2^{Q}$ different combinations for the antenna
coders and each antenna coder produces distinct radiation patterns.
As a result, we can select the antenna coder $\mathbf{b}$ among a
wide range of $2^{Q}$ combinations to find the optimum radiation
pattern for wireless systems. In the following sections, we will introduce
SISO and MIMO pixel antenna systems and show the antenna coding design
for enhancing wireless systems.

\section{SISO Pixel Antenna System}

As a starting point, we first introduce a SISO pixel antenna system,
which is the simplest form for utilizing pixel antennas, to demonstrate
the beamspace channel model, the antenna coding design, and codebook
design for antenna coding.

\subsection{Beamspace Channel Model}

The SISO pixel antenna system is equipped with a conventional antenna
at the transmitter and a pixel antenna at the receiver, as shown in
Fig. \ref{fig:SISO PAS schematic}. To model the channel for the SISO
pixel antenna system, we use the beamspace channel representation
\cite{johnson1993dudgeon}, \cite{Kalis2014}. Specifically, utilizing
a virtual channel representation, we can represent the beamspace channel
model in the angular domain, written as 
\begin{equation}
h\left(\mathbf{b}_{\textrm{R}}\right)=\mathbf{e}_{\textrm{R}}^{T}\left(\mathbf{b}_{\textrm{R}}\right)\mathbf{H}_{\textrm{V}}\mathbf{e}_{\textrm{T}}\label{eq:SISO beamspace Channel}
\end{equation}
where $\mathbf{e}_{\textrm{T}}$ is the normalized radiation pattern
of the conventional antenna at the transmitter satisfying $\left\Vert \mathbf{e}_{\textrm{T}}\right\Vert =1$,
$\mathbf{e}_{\textrm{R}}\left(\mathbf{b}_{\textrm{R}}\right)$ is
the normalized radiation pattern of the pixel antenna coded by antenna
coder $\mathbf{b}_{\textrm{R}}$ at the receiver satisfying $\left\Vert \mathbf{e}_{\textrm{R}}\left(\mathbf{b}_{\textrm{R}}\right)\right\Vert =1$,
and $\mathbf{H}_{\textrm{V}}\in\mathbb{C}^{2K\times2K}$ is the virtual
channel matrix given by
\begin{equation}
\mathbf{H}_{\textrm{V}}=\left[\begin{array}{cc}
\mathbf{H}_{\textrm{V},\theta\theta} & \mathbf{H}_{\textrm{V},\theta\phi}\\
\mathbf{H}_{\textrm{V},\phi\theta} & \mathbf{H}_{\textrm{V},\phi\phi}
\end{array}\right],\label{eq:virtual channel matrix}
\end{equation}
where $\mathbf{H}_{\textrm{V},\theta\theta}$, $\mathbf{H}_{\textrm{V},\theta\phi}$,
$\mathbf{H}_{\textrm{V},\phi\theta}$, $\mathbf{H}_{\textrm{V},\phi\phi}\in\mathbb{C}^{K\times K}$
are the virtual channel matrices for $\theta$ and $\phi$ polarizations,
respectively, with each entry being the channel gain from an angle
of departure (AoD) to an angle of arrival (AoA) among the $K$ spatial
angles. We consider a rich scattering environment with Rayleigh fading
and assume that $\left[\mathbf{H}_{\textrm{V}}\right]_{i,j}$ $\forall i,j$
are independent and identically distributed (i.i.d.) random variables
following the complex Gaussian distribution $\mathcal{CN}\left(0,1\right)$.

From the beamspace channel model \eqref{eq:SISO beamspace Channel},
we can see that the channel of SISO pixel antenna system $h\left(\mathbf{b}_{\textrm{R}}\right)$
can be coded by the antenna coder $\mathbf{b}_{\textrm{R}}$, which
allows the antenna coding optimization to control the channel and
enhance the system.

\subsection{Antenna Coding Design}

For the SISO pixel antenna system, the received signal $y\in\mathbb{C}$
can be expressed as
\begin{equation}
y=h\left(\mathbf{b}_{\textrm{R}}\right)x+n,\label{eq:SISO Model}
\end{equation}
where $x\in\mathbb{C}$ is the transmit signal and $n\in\mathbb{C}$
is additive white Gaussian noise following the complex Gaussian distribution
$\mathcal{CN}\left(0,\sigma^{2}\right)$. Leveraging \eqref{eq:SISO beamspace Channel}
and \eqref{eq:SISO Model}, the channel gain for the SISO pixel antenna
system is given by
\begin{equation}
\left|h\left(\mathbf{b}_{\textrm{R}}\right)\right|^{2}=\left|\mathbf{e}_{\textrm{R}}^{T}\left(\mathbf{b}_{\textrm{R}}\right)\mathbf{H}_{\textrm{V}}\mathbf{e}_{\textrm{T}}\right|^{2}.
\end{equation}
Assuming the channel state information (CSI) for the virtual channel
matrix $\mathbf{H}_{\textrm{V}}$ is perfectly known as the CSI can
be estimated by using methods in \cite{zhang2023successive}, we aim
to optimize the antenna coder for the pixel antenna $\mathbf{b}_{\textrm{R}}$
to maximize the channel gain, formulated as
\begin{align}
\underset{\mathbf{e}_{\textrm{R}},\mathbf{b}_{\textrm{R}},\mathbf{i}_{\textrm{R}},i_{\textrm{A}}^{\textrm{R}}}{\text{max}} & \left|\mathbf{e}_{\textrm{R}}^{T}\left(\mathbf{b}_{\textrm{R}}\right)\mathbf{H}_{\textrm{V}}\mathbf{e}_{\textrm{T}}\right|^{2}\label{eq: channel gain 0}\\
\text{s.t.}\,\ \ \  & \left\Vert \mathbf{e}_{\textrm{R}}\left(\mathbf{b}_{\textrm{R}}\right)\right\Vert =1,\label{eq: norm erbr 0}\\
 & \mathbf{e}_{\textrm{R}}\left(\mathbf{b}_{\textrm{R}}\right)=\mathbf{E}_{\textrm{oc}}\mathbf{i}_{\textrm{R}}\left(\mathbf{b}_{\textrm{R}}\right),\\
 & \mathbf{i}_{\textrm{R}}\left(\mathbf{b}_{\textrm{R}}\right)=\left[\begin{array}{c}
1\\
-\left(\mathbf{Z}_{\textrm{PP}}+\mathbf{Z}_{\textrm{L}}\left(\mathbf{b}_{\textrm{R}}\right)\right)^{-1}\mathbf{z}_{\textrm{PA}}
\end{array}\right]i_{\textrm{A}}^{\textrm{R}},\\
 & \left[\mathbf{b}_{\textrm{R}}\right]_{q}\in\left\{ 0,1\right\} ,\forall q,\label{eq: br 0}
\end{align}
where $\mathbf{i}_{\textrm{R}}\left(\mathbf{b}_{\textrm{R}}\right)$
is the currents at all ports of the pixel antenna coded by $\mathbf{b}_{\textrm{R}}$
at the receiver and $i_{\textrm{A}}^{\textrm{R}}$ is the current
at the antenna port. $i_{\textrm{A}}^{\textrm{R}}$ is an optimization
variable to ensure that the constraint \eqref{eq: norm erbr 0} can
be satisfied for any given $\mathbf{b}_{\textrm{R}}$.

To handle the constraint \eqref{eq: norm erbr 0}, we normalize the
radiation pattern $\mathbf{e}_{\textrm{R}}\left(\mathbf{b}_{\textrm{R}}\right)$
and therefore equivalently transform the problem \eqref{eq: channel gain 0}-\eqref{eq: br 0}
as
\begin{align}
\underset{\mathbf{e}_{\textrm{R}},\mathbf{b}_{\textrm{R}},\bar{\mathbf{i}}_{\textrm{R}}}{\text{max}\ \ } & \left|\mathbf{e}_{\textrm{R}}^{T}\left(\mathbf{b}_{\textrm{R}}\right)\mathbf{H}_{\textrm{V}}\mathbf{e}_{\textrm{T}}\right|^{2}\label{eq: channel gain 1}\\
\text{s.t.}\ \ \  & \mathbf{e}_{\textrm{R}}\left(\mathbf{b}_{\textrm{R}}\right)=\frac{\mathbf{E}_{\textrm{oc}}\bar{\mathbf{i}}_{\textrm{R}}\left(\mathbf{b}_{\textrm{R}}\right)}{\left\Vert \mathbf{E}_{\textrm{oc}}\bar{\mathbf{i}}_{\textrm{R}}\left(\mathbf{b}_{\textrm{R}}\right)\right\Vert },\label{eq: norm pattern}\\
 & \bar{\mathbf{i}}_{\textrm{R}}\left(\mathbf{b}_{\textrm{R}}\right)=\left[\begin{array}{c}
1\\
-\left(\mathbf{Z}_{\textrm{PP}}+\mathbf{Z}_{\textrm{L}}\left(\mathbf{b}_{\textrm{R}}\right)\right)^{-1}\mathbf{z}_{\textrm{PA}}
\end{array}\right],\label{eq: norm current}\\
 & \left[\mathbf{b}_{\textrm{R}}\right]_{q}\in\left\{ 0,1\right\} ,\forall q,\label{eq: br 0 0}
\end{align}
where the variable $i_{\textrm{A}}^{\textrm{R}}$ is removed because
it does not affect the normalized radiation pattern. Substituting
\eqref{eq: norm current} into \eqref{eq: norm pattern}, we can find
the expression of $\mathbf{e}_{\textrm{R}}\left(\mathbf{b}_{\textrm{R}}\right)$
as a function of $\mathbf{b}_{\textrm{R}}$ and thus equivalently
simplify the problem \eqref{eq: channel gain 1}-\eqref{eq: br 0 0}
as
\begin{align}
\underset{\mathbf{b}_{\textrm{R}}}{\text{max}\ \ } & \left|\mathbf{e}_{\textrm{R}}^{T}\left(\mathbf{b}_{\textrm{R}}\right)\mathbf{H}_{\textrm{V}}\mathbf{e}_{\textrm{T}}\right|^{2}\label{eq:objective channel gain}\\
\text{s.t.}\ \ \  & \left[\mathbf{b}_{\textrm{R}}\right]_{q}\in\left\{ 0,1\right\} ,\forall q,\label{eq:bR 0 1}
\end{align}
which is an NP-hard binary optimization problem.

Various algorithms have been proposed and utilized for pixel antenna
design and optimization such as the genetic algorithm \cite{Pixel_Song},
mixed integer linear programming \cite{ShanpuShen2017_APS_MILP},
N-port characteristic mode analysis \cite{8952871}, \cite{9810821},
perturbation sensitivity analysis \cite{9491941}, the adjoint method
combined with the method of moving asymptote \cite{10035928}, and
reinforcement learning \cite{Chen2024}. In this work, to solve the
problem \eqref{eq:objective channel gain}-\eqref{eq:bR 0 1}, we
use an efficient algorithm called the Successive Exhaustive Boolean
Optimization (SEBO) \cite{ShanpuShen2017_TAP_SEBO}, which has two
steps: 1) cyclically optimizing each block of the binary variables
by exhaustive search until convergence, and 2) randomly flipping bits
in the converged binary solution to check if there is any other local
optimum with a better objective value. The computational complexity
of SEBO is $\mathcal{O}\left(N_{\textrm{e}}2^{J}\right)$ where $J$
is the block size and $N_{\textrm{e}}$ is the number of iterations.
Increasing $J$ can enhance the optimization performance. Particularly,
when $J=Q$, the SEBO becomes the exhaustive search and global optimal
solution can be found. However, increasing $J$ will exponentially
increase the computational complexity. Thus, there is a trade-off
between the computational complexity and optimization performance
when using the SEBO to optimize antenna coding. More details on SEBO
can be found in \cite{ShanpuShen2017_TAP_SEBO}.

\subsection{Codebook Design for Antenna Coding}

To reduce the computational complexity, we design a codebook for antenna
coding optimization. Given a codebook which is defined as a set of
$M$ different antenna coders $\mathcal{B}\triangleq\left\{ \mathbf{b}_{1},\mathbf{b}_{2},\ldots,\mathbf{b}_{M}\right\} $,
the channel gain is maximized by selecting the antenna coder from
the codebook, formulated as
\begin{equation}
\mathbf{b}_{\textrm{R}}^{\star}=\underset{\mathbf{b}_{\textrm{R}}\in\mathcal{B}}{\text{argmax}}\left|\mathbf{e}_{\textrm{R}}^{T}\left(\mathbf{b}_{\textrm{R}}\right)\mathbf{H}_{\textrm{V}}\mathbf{e}_{\textrm{T}}\right|^{2}.\label{eq:OPTIMIZE WITH CDB}
\end{equation}
Given a codebook and channel realization of $\mathbf{H}_{\textrm{V}}$,
we can optimize the antenna coder $\mathbf{b}_{\textrm{R}}$ by searching
the codebook. However, it is important to design a good codebook to
maximize ergodic channel gain, formulated as
\begin{align}
\underset{\mathbf{b}_{m}}{\text{max}}\ \  & \mathbb{E}\left[\left|\mathbf{e}_{\textrm{R}}^{T}\left(\mathbf{b}_{\textrm{R}}^{\star}\right)\mathbf{H}_{\textrm{V}}\mathbf{e}_{\textrm{T}}\right|^{2}\right]\label{eq:codebook design problem 1}\\
\text{s.t.}\ \ \  & \left[\mathbf{b}_{m}\right]_{q}\in\left\{ 0,1\right\} ,\forall m,q.\label{eq:codebook design problem 2}
\end{align}

The codebook design problem \eqref{eq:codebook design problem 1}
and \eqref{eq:codebook design problem 2} is related to the vector
quantization problem \cite{gersho2012vector}. To handle the ergodic
channel gain in \eqref{eq:codebook design problem 1}, we leverage
the sample average approximation method \cite{Shapiro2021}. Specifically,
we consider a training set containing $L$ realizations for the virtual
channel matrix $\mathbf{H}_{\textrm{V}}$ which is a random matrix
with each entry following i.i.d. complex Gaussian distribution, given
by
\begin{equation}
\mathcal{H}\triangleq\left\{ \mathbf{H}_{\textrm{V}}^{\left[l\right]}\mid l=1,\ldots,L\right\} ,
\end{equation}
where $\mathbf{H}_{\textrm{V}}^{\left[l\right]}$ denotes the $l$th
virtual channel realization. Accordingly, we can approximate the expectation
of channel gain in \eqref{eq:codebook design problem 1} as the sample
average of channel gain, written as
\begin{equation}
\mathbb{E}\left[\left|\mathbf{e}_{\textrm{R}}^{T}\left(\mathbf{b}_{\textrm{R}}^{\star}\right)\mathbf{H}_{\textrm{V}}\mathbf{e}_{\textrm{T}}\right|^{2}\right]\approx\frac{1}{L}\sum_{l=1}^{L}\left|\mathbf{e}_{\textrm{R}}^{T}\left(\mathbf{b}_{\textrm{R}}^{\star}\right)\mathbf{H}_{\textrm{V}}^{\left[l\right]}\mathbf{e}_{\textrm{T}}\right|^{2}.\label{eq:SAA}
\end{equation}
Associated with each antenna coder, we can partition the training
set $\mathcal{H}$ into $M$ subsets $\mathcal{H}_{1}$, $\mathcal{H}_{2}$,
$\dots$, $\mathcal{H}_{M}$, where $\mathcal{H}_{m}$ is the neighborhood
of the antenna coder $\mathbf{b}_{m}$ defined as
\begin{align}
\mathcal{H}_{m} & =\left\{ \mathbf{H}_{\textrm{V}}^{\left[l\right]}\mid\left|\mathbf{e}_{\textrm{R}}^{T}\left(\mathbf{b}_{m}\right)\mathbf{H}_{\textrm{V}}^{\left[l\right]}\mathbf{e}_{\textrm{T}}\right|^{2}\right.\nonumber \\
 & \left.\qquad\hfill\geq\left|\mathbf{e}_{\textrm{R}}^{T}\left(\mathbf{b}_{m^{\prime}}\right)\mathbf{H}_{\textrm{V}}^{\left[l\right]}\mathbf{e}_{\textrm{T}}\right|^{2},\forall l,\forall m^{\prime}\neq m\right\} ,\forall m.\label{eq:partition definition}
\end{align}
Making use of \eqref{eq:SAA} and \eqref{eq:partition definition},
we can equivalently reformulate the problem \eqref{eq:codebook design problem 1}
and \eqref{eq:codebook design problem 2} as
\begin{align}
\underset{\mathbf{b}_{m}}{\text{max}}\ \  & \sum_{m=1}^{M}\sum_{\mathbf{H}_{\textrm{V}}^{\left[l\right]}\in\mathcal{H}_{m}}\left|\mathbf{e}_{\textrm{R}}^{T}\left(\mathbf{b}_{m}\right)\mathbf{H}_{\textrm{V}}^{\left[l\right]}\mathbf{e}_{\textrm{T}}\right|^{2}\label{eq:codebook design problem SAA 1}\\
\text{s.t.}\ \ \  & \left[\mathbf{b}_{m}\right]_{q}\in\left\{ 0,1\right\} ,\forall m,q.\label{eq:codebook design problem SAA 2}
\end{align}
Note that to ensure that the sample average accurately approximates
the ergodic channel gain, the entries of training set should be the
independent realizations of random $\mathbf{H}_{\textrm{V}}$ and
the size of training set $L$ should be large enough compared to the
codebook size. Otherwise, such approximation is inaccurate and will
degrade the codebook design.

We can see that the partition of training set $\mathcal{H}_{m}$ $\forall m$
relies on the antenna coder $\mathbf{b}_{m}$ $\forall m$ in the
problem \eqref{eq:codebook design problem SAA 1} and \eqref{eq:codebook design problem SAA 2}.
Such coupled $\mathcal{H}_{m}$ and $\mathbf{b}_{m}$ make the problem
\eqref{eq:codebook design problem SAA 1} and \eqref{eq:codebook design problem SAA 2}
intractable. To address the problem \eqref{eq:codebook design problem SAA 1}
and \eqref{eq:codebook design problem SAA 2}, we follow the framework
of the generalized Lloyd algorithm (GLA) \cite{xia2006design} to
optimize the training set partition $\mathcal{H}_{m}$ and antenna
coder $\mathbf{b}_{m}$ alternatively via leveraging the nearest neighbor
rule and the centroid condition. The methods are described next.

\subsubsection{Partition Optimization}

One necessary condition for the optimal codebook is the nearest neighbor
rule. Namely, all the channel realizations which have higher channel
gain with the antenna coder $\mathbf{b}_{m}$ than with any other
antenna coders should be assigned to the partition $\mathcal{H}_{m}$.
Therefore, at iteration $i$ of the alternating optimization, we first
optimize the partition by
\begin{align}
\mathcal{H}_{m}^{\left(i\right)} & =\left\{ \mathbf{H}_{\textrm{V}}^{\left[l\right]}\mid\left|\mathbf{e}_{\textrm{R}}^{T}\left(\mathbf{b}_{m}^{\left(i-1\right)}\right)\mathbf{H}_{\textrm{V}}^{\left[l\right]}\mathbf{e}_{\textrm{T}}\right|^{2}\right.\nonumber \\
 & \left.\qquad\hfill\geq\left|\mathbf{e}_{\textrm{R}}^{T}\left(\mathbf{b}_{m^{\prime}}^{\left(i-1\right)}\right)\mathbf{H}_{\textrm{V}}^{\left[l\right]}\mathbf{e}_{\textrm{T}}\right|^{2},\forall l,\forall m^{\prime}\neq m\right\} ,\forall m.\label{eq:partition optimization}
\end{align}
where $\mathbf{b}_{m}^{\left(i-1\right)}$ is the antenna coder optimized
at iteration $i-1$.

\subsubsection{Antenna Coder Optimization}

The other necessary condition for the optimal codebook is the centroid
condition. Namely, the optimal antenna coder $\mathbf{b}_{m}$ should
be selected to maximize the average channel gain over the partition
$\mathcal{H}_{m}$. Therefore, at iteration $i$, after the partition
optimization, we then optimize the antenna coder by
\begin{align}
\mathbf{b}_{m}^{\left(i\right)} & =\underset{\left[\mathbf{b}_{m}\right]_{q}\in\left\{ 0,1\right\} ,\forall m,q}{\textrm{argmax}}\sum_{\mathbf{H}_{\textrm{V}}^{\left[l\right]}\in\mathcal{H}_{m}^{\left(i\right)}}\left|\mathbf{e}_{\textrm{R}}^{T}\left(\mathbf{b}_{m}\right)\mathbf{H}_{\textrm{V}}^{\left[l\right]}\mathbf{e}_{\textrm{T}}\right|^{2},\forall m,\label{eq:codeword optimization}
\end{align}
which is a binary optimization problem that can be solved by the SEBO
algorithm.

By alternatively optimizing the partition \eqref{eq:partition optimization}
and the antenna coder \eqref{eq:codeword optimization}, the average
channel gain increases over each iteration so that the codebook design
converges. Algorithm \ref{alg:Codebook-Design-} summarizes the overall
algorithm for codebook design.

\begin{algorithm}[t]
\caption{\label{alg:Codebook-Design-}Codebook Design for Antenna Coding}

\floatname{algorithm}{Algorithm}   
\renewcommand{\algorithmicrequire}{\textbf{Input:}} 
\renewcommand{\algorithmicensure}{\textbf{Output:}}  
\newcommand{\INDSTATE}[1][1]{\STATE\hspace{#1\algorithmicindent}}                                         
\begin{algorithmic}[1]                       
\REQUIRE             $\mathcal{H}$;                                   
\ENSURE              Codebook design $\mathcal{B}$;                      
\STATE               \textbf{Initialization:} $i=0$, $\mathbf{b}_{m}^{\left(0\right)}$ $\forall m$;                                              
\STATE               \textbf{repeat} \\  
                     $\text{\ \ \ \ \ \ } i=i+1;$                           
\STATE               $\text{\ \ \ \ \ \ } \textbf{Partition\ Optimization:}$ \\ 
                     $\text{\ \ \ \ \ \ } \text{Update} \ \mathcal{H}_{m}^{\left(i\right)} \ \forall m$ by (\ref{eq:partition optimization}) with $\mathbf{b}_{m}^{\left(i-1\right)}$;\\                                                    
\STATE               $\text{\ \ \ \ \  }\ \textbf{Antenna Coder\ Optimization:}$\\
                     $\text{\ \ \ \ \ \ } \text{Update} \ \mathbf{b}_{m}^{\left(i\right)} \ \forall m$  by solving (\ref{eq:codeword optimization}) with $\mathcal{H}_{m}^{\left(i\right)}$;\\                                                       \STATE              \textbf{until} $\sum\left\Vert \mathbf{b}_{m}^{\left(i\right)}-\mathbf{b}_{m}^{\left(i-1\right)}\right\Vert /\sum\left\Vert \mathbf{b}_{m}^{\left(i\right)}\right\Vert \leq\epsilon$ or  $i=i_{\textrm{max}}$;                           
\STATE               Obtain $\mathbf{b}_{m}=\mathbf{b}_{m}^{\left(i\right)}$  $\forall m$;                                   
\STATE               Obtain $\mathcal{B}\triangleq\left\{ \mathbf{b}_{1},\mathbf{b}_{2},\ldots,\mathbf{b}_{M}\right\}$                                                                                             \end{algorithmic}
\end{algorithm}

\subsection{Performance Analysis}

\begin{figure*}[tbh]
\begin{centering}
\includegraphics[width=17.5cm]{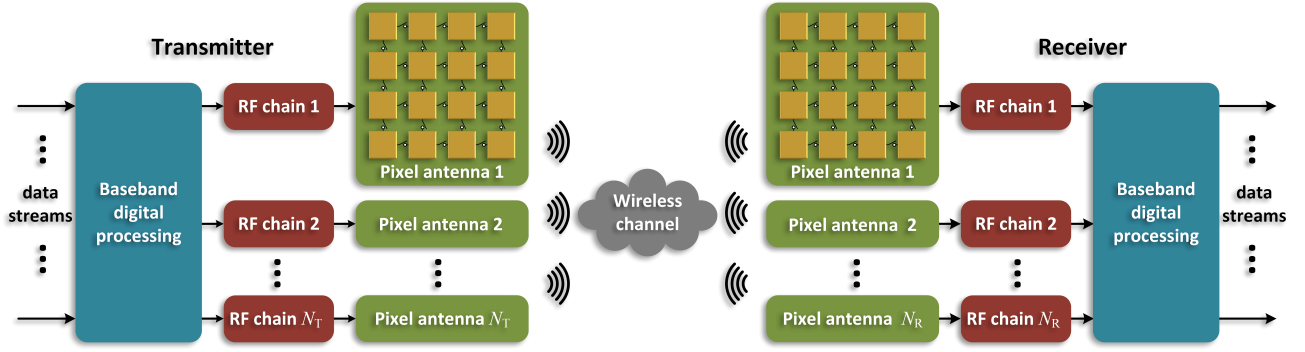}
\par\end{centering}
\caption{\label{fig:MIMO PAS schematic}System diagram of the MIMO pixel antenna
system.}
\end{figure*}

In order to obtain insights into the fundamental limits of pixel antennas,
we derive the upper bound of the channel gain of SISO pixel antenna
system as follows.

Using singular value decomposition (SVD), we can decompose the open-circuit
radiation pattern matrix $\mathbf{E}_{\textrm{oc}}$ as
\begin{equation}
\mathbf{E}_{\textrm{oc}}=\mathbf{U}\mathbf{S}\mathbf{V}^{H}\label{eq:SVD}
\end{equation}
where $R=\textrm{rank}\left(\mathbf{E}_{\textrm{oc}}\right)$ is the
rank of matrix $\mathbf{E}_{\textrm{oc}}$, $\mathbf{U}\in\mathbb{C}^{2K\times R}$
and $\mathbf{V}\in\mathbb{C}^{(Q+1)\times R}$ are semi-unitary matrices
satisfying $\mathbf{U}^{H}\mathbf{U}=\mathbf{V}^{H}\mathbf{V}=\mathbf{I}$,
and $\mathbf{S}=\textrm{diag}\left(s_{1},\ldots,s_{R}\right)\in\mathbb{R}^{R\times R}$
with $s_{i}$ representing the $i$th singular value. Making use of
\eqref{eq:F pattern} and \eqref{eq:SVD}, we can rewrite $\mathbf{e}_{\textrm{R}}\left(\mathbf{b}_{\textrm{R}}\right)$
as
\begin{equation}
\mathbf{e}_{\textrm{R}}\left(\mathbf{b}_{\textrm{R}}\right)=\mathbf{E}_{\textrm{oc}}\mathbf{i}_{\textrm{R}}\left(\mathbf{b}_{\textrm{R}}\right)=\mathbf{U}\mathbf{S}\mathbf{V}^{H}\mathbf{i}_{\textrm{R}}\left(\mathbf{b}_{\textrm{R}}\right).\label{eq:er SVD}
\end{equation}
This shows that any possible radiation pattern generated by the pixel
antenna can be decomposed into $R$ orthogonal radiation patterns,
i.e. the $R$ columns of $\mathbf{U}$. In the beamspace domain, the
number of orthogonal radiation patterns that antennas can provide
is defined as effective aerial degrees-of-freedom (EADoF) \cite{9286866}.
Hence, the EADoF of pixel antennas is given by $R$. Moreover, the
singular value $s_{i}$ reflects the ability to radiate for the $i$th
orthogonal radiation pattern. In other words, $s_{i}^{2}$ characterizes
how much the power of the $i$th orthogonal radiation pattern is in
the total radiated power. Thus, a high value of $s_{i}^{2}$ implies
that the $i$th orthogonal radiation pattern is dominant in the total
radiation pattern of the pixel antenna.

Accordingly, substituting \eqref{eq:er SVD} into \eqref{eq:SISO beamspace Channel},
we can rewrite the beamspace channel of the SISO pixel antenna system
as
\begin{equation}
h\left(\mathbf{b}_{\textrm{R}}\right)=\mathbf{i}_{\textrm{R}}^{T}\left(\mathbf{b}_{\textrm{R}}\right)\mathbf{V}^{*}\mathbf{S}\mathbf{U}^{T}\mathbf{H}_{\textrm{V}}\mathbf{e}_{\textrm{T}}=\mathbf{w}^{H}\left(\mathbf{b}_{\textrm{R}}\right)\widetilde{\mathbf{h}},\label{eq:SIMO CHANNEL}
\end{equation}
where $\mathbf{w}\left(\mathbf{b}_{\textrm{R}}\right)\in\mathbb{C}^{R\times1}$
is defined as
\begin{equation}
\mathbf{w}\left(\mathbf{b}_{\textrm{R}}\right)=\mathbf{S}\mathbf{V}^{T}\mathbf{i}_{\textrm{R}}^{*}\left(\mathbf{b}_{\textrm{R}}\right),\label{eq:equivalent combiner}
\end{equation}
which satisfies $\left\Vert \mathbf{w}\left(\mathbf{b}_{\textrm{R}}\right)\right\Vert =1$
due to the normalized radiation pattern $\left\Vert \mathbf{e}_{\textrm{R}}\left(\mathbf{b}_{\textrm{R}}\right)\right\Vert =1$,
and $\widetilde{\mathbf{h}}\in\mathbb{C}^{R\times1}$ is defined as
\begin{equation}
\widetilde{\mathbf{h}}=\mathbf{U}^{T}\mathbf{H}_{\textrm{V}}\mathbf{e}_{\textrm{T}},
\end{equation}
where $[\widetilde{\mathbf{h}}]_{i}$ $\forall i$ are i.i.d. random
variables following the complex Gaussian distribution $\mathcal{CN}\left(0,1\right)$
because of the $R$ orthogonal radiation patterns collected in $\mathbf{U}$.

From \eqref{eq:SIMO CHANNEL}, we can observe that the channel of
SISO pixel antenna system (with a conventional antenna at the transmitter
and a pixel antenna at the receiver) is equivalent to a single-input
multiple-output (SIMO) channel with combining. Therefore, we can find
an upper bound for the channel gain of SISO pixel antenna system using
Cauchy-Schwarz inequality
\begin{equation}
\left|h\left(\mathbf{b}_{\textrm{R}}\right)\right|^{2}=\left|\mathbf{w}^{H}\left(\mathbf{b}_{\textrm{R}}\right)\widetilde{\mathbf{h}}\right|^{2}\leq\left\Vert \widetilde{\mathbf{h}}\right\Vert ^{2},
\end{equation}
where the upper bound is achieved if and only if the maximum ratio
combiner is achieved, i.e.
\begin{equation}
\mathbf{w}\left(\mathbf{b}_{\textrm{R}}\right)=\frac{\widetilde{\mathbf{h}}}{\left\Vert \widetilde{\mathbf{h}}\right\Vert }.\label{eq:MRC}
\end{equation}
However, it should be noted that for pixel antennas there is no guarantee
that an antenna coder $\mathbf{b}_{\textrm{R}}$, such that $\mathbf{w}\left(\mathbf{b}_{\textrm{R}}\right)$
satisfies \eqref{eq:MRC}, can be found. Therefore the upper bound
for channel gain can be approached but cannot be guaranteed to be
achieved. Accordingly, the upper bound for the average channel gain
of SISO pixel antenna system is given by
\begin{equation}
\mathbb{E}\left[\left|h\left(\mathbf{b}_{\textrm{R}}\right)\right|^{2}\right]\leq\mathbb{E}\left[\left\Vert \widetilde{\mathbf{h}}\right\Vert ^{2}\right]=R.\label{eq:upper bound}
\end{equation}
Therefore, this indicates that the average channel gain of the SISO
pixel antenna system is upper bounded by how many orthogonal radiation
patterns the pixel antenna can generate, i.e. the EADoF of pixel antennas,
which is determined by the physical aperture of the pixel antenna\footnote{It is difficult to write the relationship between EADoF and aperture
in closed form as $\mathbf{E}_{\textrm{oc}}$ is obtained by simulating
the pixel antenna by EM solver.}. The average channel gain in comparison with the corresponding upper
bound for SISO pixel antenna system with different physical apertures
will be shown in Section V.

\section{MIMO Pixel Antenna System}

We introduce a MIMO pixel antenna system to demonstrate the beamspace
channel model and antenna coding design for capacity maximization
in this section.

\subsection{Beamspace Channel Model}

We consider a MIMO pixel antenna system. As shown in Fig. \ref{fig:MIMO PAS schematic},
the transmitter is equipped with $N_{\textrm{T}}$ pixel antennas,
each of which is coded by $\mathbf{b}_{\textrm{T},n_{\textrm{T}}}\,\forall n_{\textrm{T}}$,
and the receiver is equipped with $N_{\textrm{R}}$ pixel antennas,
each of which is coded by $\mathbf{b}_{\textrm{R},n_{\textrm{R}}}\,\forall n_{\textrm{R}}$.
We group $\mathbf{b}_{\textrm{T},n_{\textrm{T}}}\,\forall n_{\textrm{T}}$
into a matrix $\mathbf{B}_{\textrm{T}}=\left[\mathbf{b}_{\textrm{T},1},\ldots,\mathbf{b}_{\textrm{T},N_{\textrm{T}}}\right]\in\mathbb{R}^{Q\times N_{\textrm{T}}}$
and $\mathbf{b}_{\textrm{R},n_{\textrm{R}}}\,\forall n_{\textrm{R}}$
into a matrix $\mathbf{B}_{\textrm{R}}=\left[\mathbf{b}_{\textrm{R},1},\ldots,\mathbf{b}_{\textrm{R},N_{\textrm{R}}}\right]\in\mathbb{R}^{Q\times N_{\textrm{R}}}$.
Utilizing the beamspace channel representation, the channel for MIMO
pixel antenna system can be expressed as
\begin{equation}
\mathbf{H}\left(\mathbf{B}_{\textrm{T}},\mathbf{B}_{\textrm{R}}\right)=\mathbf{E}_{\textrm{R}}^{T}\left(\mathbf{B}_{\textrm{R}}\right)\mathbf{H}_{\textrm{V}}\mathbf{E}_{\textrm{T}}\left(\mathbf{B}_{\textrm{T}}\right)\label{eq:beamspace MIMO pixel antenna}
\end{equation}
where $\mathbf{E}_{\textrm{T}}\left(\mathbf{B}_{\textrm{T}}\right)\triangleq\text{[\ensuremath{\mathbf{e}_{\textrm{T},1}\left(\mathbf{b}_{\textrm{T},1}\right)},\ensuremath{\ldots},\ensuremath{\mathbf{e}_{\textrm{T},N_{\textrm{T}}}\left(\mathbf{b}_{\textrm{T},N_{\textrm{T}}}\right)}]}\in\mathbb{C}^{2K\times N_{\textrm{T}}}$
with $\mathbf{e}_{\textrm{T},n_{\textrm{T}}}\left(\mathbf{b}_{\textrm{T},n_{\textrm{T}}}\right)$
being the normalized radiation pattern of the $n_{\textrm{T}}$th
transmit pixel antenna satisfying $\left\Vert \mathbf{e}_{\textrm{T},n_{\textrm{T}}}\left(\mathbf{b}_{\textrm{T},n_{\textrm{T}}}\right)\right\Vert =1$,
$\mathbf{E}_{\textrm{R}}\left(\mathbf{B}_{\textrm{R}}\right)\triangleq[\mathbf{e}_{\textrm{R},1}\left(\mathbf{b}_{\textrm{R},1}\right),\ldots,\mathbf{e}_{\textrm{R},N_{\textrm{R}}}\left(\mathbf{b}_{\textrm{R},N_{\textrm{R}}}\right)]\in\mathbb{C}^{2K\times N_{\textrm{R}}}$
with $\mathbf{e}_{\textrm{R},n_{\textrm{R}}}\left(\mathbf{b}_{\textrm{R},n_{\textrm{R}}}\right)$
being the normalized radiation pattern of the $n_{\textrm{R}}$th
receive pixel antenna satisfying $\left\Vert \mathbf{e}_{\textrm{R},n_{\textrm{R}}}\left(\mathbf{b}_{\textrm{R},n_{\textrm{R}}}\right)\right\Vert =1$,
and $\mathbf{H}_{\textrm{V}}\in\mathbb{C}^{2K\times2K}$ is the virtual
channel matrix given by \eqref{eq:virtual channel matrix}. Similarly,
from the beamspace channel model \eqref{eq:beamspace MIMO pixel antenna},
we can see that the channel of MIMO pixel antenna system $\mathbf{H}\left(\mathbf{B}_{\textrm{T}},\mathbf{B}_{\textrm{R}}\right)$
can be coded by the transmit and receive antenna coders $\mathbf{B}_{\textrm{T}}$
and $\mathbf{B}_{\textrm{R}}$, which allows antenna coding optimization
to enhance the MIMO system.

\subsection{Channel Capacity}

For the MIMO pixel antenna system, the received signal $\mathbf{y}\in\mathbb{C}^{N_{\textrm{R}}\times1}$
can be expressed as
\begin{equation}
\mathbf{y}=\mathbf{H}\left(\mathbf{B}_{\textrm{T}},\mathbf{B}_{\textrm{R}}\right)\mathbf{x}+\mathbf{n},
\end{equation}
where $\mathbf{x}\in\mathbb{C}^{N_{\textrm{T}}\times1}$ is the transmit
signal with covariance matrix given by $\mathbf{X}=\mathbb{E}\left[\mathbf{x}\mathbf{x}^{H}\right]$
and $\mathbf{n}\in\mathbb{C}^{N_{\textrm{R}}\times1}$ is the additive
white Gaussian noise following the complex Gaussian distribution $\mathcal{CN}\left(\boldsymbol{0},\sigma^{2}\mathbf{I}\right)$.
Accordingly, the channel capacity for the MIMO pixel antenna system
is given by
\begin{equation}
C\left(\mathbf{X},\mathbf{B}_{\textrm{T}},\mathbf{B}_{\textrm{R}}\right)=\log_{2}\left|\mathbf{I}+\frac{1}{\sigma^{2}}\mathbf{H}\left(\mathbf{B}_{\textrm{T}},\mathbf{B}_{\textrm{R}}\right)\mathbf{X}\mathbf{H}^{H}\left(\mathbf{B}_{\textrm{T}},\mathbf{B}_{\textrm{R}}\right)\right|.
\end{equation}
Assuming the CSI for the virtual channel matrix $\mathbf{H}_{\textrm{V}}$
is perfectly known, we aim to jointly optimize the transmit signal
covariance matrix $\mathbf{X}$ and the transmit and receive antenna
coders $\mathbf{B}_{\textrm{T}}$ and $\mathbf{B}_{\textrm{R}}$ to
maximize the channel capacity for the MIMO pixel antenna system, which
can be formulated as
\begin{align}
\underset{\mathbf{X},\mathbf{B}_{\textrm{T}},\mathbf{B}_{\textrm{R}}}{\text{max}} & \log_{2}\left|\mathbf{I}+\frac{1}{\sigma^{2}}\mathbf{H}\left(\mathbf{B}_{\textrm{T}},\mathbf{B}_{\textrm{R}}\right)\mathbf{X}\mathbf{H}^{H}\left(\mathbf{B}_{\textrm{T}},\mathbf{B}_{\textrm{R}}\right)\right|\label{eq:objective capacity maximization}\\
\text{s.t.}\ \ \  & \textrm{Tr}\left(\mathbf{X}\right)\leq P,\label{eq:constraint TrX<P}\\
 & \left[\mathbf{B}_{\textrm{T}}\right]_{i,j}\in\left\{ 0,1\right\} ,\forall i,j,\label{eq:BT=00003D0,1 General}\\
 & \left[\mathbf{B}_{\textrm{R}}\right]_{i,j}\in\left\{ 0,1\right\} ,\forall i,j,\label{eq:BR=00003D0,1 General}
\end{align}
where $P$ denotes the maximum transmit power.

\subsection{Antenna Coding Design}

To solve the capacity maximization problem \eqref{eq:objective capacity maximization}-\eqref{eq:BR=00003D0,1 General},
in the following, we consider two cases of the transmit signal covariance
matrix.

\subsubsection{Antenna Coding Design with Uniform Power Allocation}

We first consider when the uniform power allocation is utilized in
the MIMO pixel antenna system, i.e. $\mathbf{X}=\frac{P}{N_{\textrm{T}}}\mathbf{I}$.
In this case, we only need to optimize the transmit and receive antenna
coders $\mathbf{B}_{\textrm{T}}$ and $\mathbf{B}_{\textrm{R}}$ for
capacity maximization, formulated as
\begin{align}
\underset{\mathbf{B}_{\textrm{T}},\mathbf{B}_{\textrm{R}}}{\text{max}}\  & \log_{2}\left|\mathbf{I}+\frac{P}{\sigma^{2}N_{\textrm{T}}}\mathbf{H}\left(\mathbf{B}_{\textrm{T}},\mathbf{B}_{\textrm{R}}\right)\mathbf{H}^{H}\left(\mathbf{B}_{\textrm{T}},\mathbf{B}_{\textrm{R}}\right)\right|\label{eq:objective capacity maximization UPA}\\
\text{s.t.}\ \ \  & \left[\mathbf{B}_{\textrm{T}}\right]_{i,j}\in\left\{ 0,1\right\} ,\:\forall i,j,\label{eq:BT=00003D0,1 UPA}\\
 & \left[\mathbf{B}_{\textrm{R}}\right]_{i,j}\in\left\{ 0,1\right\} ,\:\forall i,j,\label{eq:BR=00003D0,1 UPA}
\end{align}
which is an NP-hard binary optimization problem. Similar to the channel
gain maximization in the SISO pixel antenna, we use the SEBO algorithm
to solve the problem \eqref{eq:objective capacity maximization UPA}-\eqref{eq:BR=00003D0,1 UPA}.

\subsubsection{Joint Antenna Coding and Waterfilling Design}

We next consider joint antenna coding and transmit signal covariance
matrix design for capacity maximization in the MIMO pixel antenna
system. For the channel matrix $\mathbf{H}\left(\mathbf{B}_{\textrm{T}},\mathbf{B}_{\textrm{R}}\right)$
with any given transmit and receive antenna coders $\mathbf{B}_{\textrm{T}}$
and $\mathbf{B}_{\textrm{R}}$, it is well known that the transmit
signal covariance matrix design based on waterfilling power allocation
provides the maximum channel capacity \cite{tse2005fundamentals},
i.e.
\begin{equation}
\sum_{i=1}^{N_{\textrm{min}}}\log_{2}\left(1+\frac{P_{i}^{\star}\left(\mathbf{B}_{\textrm{T}},\mathbf{B}_{\textrm{R}}\right)\lambda_{i}\left(\mathbf{B}_{\textrm{T}},\mathbf{B}_{\textrm{R}}\right)}{\sigma^{2}}\right),
\end{equation}
where $N_{\textrm{min}}=\textrm{min}\left(N_{\textrm{T}},N_{\textrm{R}}\right)$,
$\lambda_{i}\left(\mathbf{B}_{\textrm{T}},\mathbf{B}_{\textrm{R}}\right)$
is the $i$th eigenvalue of the matrix $\mathbf{H}\left(\mathbf{B}_{\textrm{T}},\mathbf{B}_{\textrm{R}}\right)\mathbf{H}^{H}\left(\mathbf{B}_{\textrm{T}},\mathbf{B}_{\textrm{R}}\right)$,
and $P_{i}^{\star}\left(\mathbf{B}_{\textrm{T}},\mathbf{B}_{\textrm{R}}\right)$
represents the waterfilling power allocation expressed as
\begin{equation}
P_{i}^{\star}\left(\mathbf{B}_{\textrm{T}},\mathbf{B}_{\textrm{R}}\right)=\left(\mu-\frac{\sigma^{2}}{\lambda_{i}\left(\mathbf{B}_{\textrm{T}},\mathbf{B}_{\textrm{R}}\right)}\right)^{+},
\end{equation}
with $\mu$ chosen to satisfy the total transmit power constraint
\begin{equation}
\sum_{i=1}^{N_{\textrm{min}}}P_{i}^{\star}\left(\mathbf{B}_{\textrm{T}},\mathbf{B}_{\textrm{R}}\right)=P.
\end{equation}
 As a result, the channel capacity with waterfilling power allocation
is also a function of $\mathbf{B}_{\textrm{T}}$ and $\mathbf{B}_{\textrm{R}}$,
so that the antenna coding design can be formulated as
\begin{align}
\underset{\mathbf{B}_{\textrm{T}},\mathbf{B}_{\textrm{R}}}{\text{max}}\  & \sum_{i=1}^{N_{\textrm{min}}}\log_{2}\left(1+\frac{P_{i}^{\star}\left(\mathbf{B}_{\textrm{T}},\mathbf{B}_{\textrm{R}}\right)\lambda_{i}\left(\mathbf{B}_{\textrm{T}},\mathbf{B}_{\textrm{R}}\right)}{\sigma^{2}}\right)\label{eq:objective capacity maximization WF}\\
\text{s.t.}\ \ \  & \left[\mathbf{B}_{\textrm{T}}\right]_{i,j}\in\left\{ 0,1\right\} ,\:\forall i,j,\label{eq:BT=00003D0,1 WF}\\
 & \left[\mathbf{B}_{\textrm{R}}\right]_{i,j}\in\left\{ 0,1\right\} ,\:\forall i,j,\label{eq:BR=00003D0,1 WF}
\end{align}
which is an NP-hard binary optimization problem. Similarly, we use
the SEBO algorithm to solve the problem \eqref{eq:objective capacity maximization WF}-\eqref{eq:BR=00003D0,1 WF}.

\subsection{Antenna Coding Design with Codebook}

To reduce computational complexity, we use the codebook proposed in
Section III-C to facilitate antenna coding design in the MIMO pixel
antenna system. Specifically, given the codebook $\mathcal{B}\triangleq\left\{ \mathbf{b}_{1},\mathbf{b}_{2},\ldots,\mathbf{b}_{M}\right\} $,
we select the antenna coder for each transmit and receive pixel antenna
from the codebook, i.e. $\mathbf{b}_{\textrm{T},n_{\textrm{T}}},\mathbf{b}_{\textrm{R},n_{\textrm{R}}}\in\mathcal{B}$
$\forall n_{\textrm{T}},n_{\textrm{R}}$, to maximize the channel
capacity of MIMO pixel antenna system. For the uniform power allocation
case, we formulate the channel capacity maximization as
\begin{align}
\underset{\mathbf{B}_{\textrm{T}},\mathbf{B}_{\textrm{R}}}{\text{max}}\  & \log_{2}\left|\mathbf{I}+\frac{P}{\sigma^{2}N_{\textrm{T}}}\mathbf{H}\left(\mathbf{B}_{\textrm{T}},\mathbf{B}_{\textrm{R}}\right)\mathbf{H}^{H}\left(\mathbf{B}_{\textrm{T}},\mathbf{B}_{\textrm{R}}\right)\right|\label{eq:objective capacity CDB UPA}\\
\text{s.t.}\ \ \  & \left[\mathbf{B}_{\textrm{T}}\right]_{:,n_{\textrm{T}}}\in\mathcal{B},\:\forall n_{\textrm{T}},\label{eq:BT CDB UPA}\\
 & \left[\mathbf{B}_{\textrm{R}}\right]_{:,n_{\textrm{R}}}\in\mathcal{B},\:\forall n_{\textrm{R}}.\label{eq:BR CDB UPA}
\end{align}
For the waterfilling power allocation case, we formulate the channel
capacity maximization as
\begin{align}
\underset{\mathbf{B}_{\textrm{T}},\mathbf{B}_{\textrm{R}}}{\text{max}}\  & \sum_{i=1}^{N_{\textrm{min}}}\log_{2}\left(1+\frac{P_{i}^{\star}\left(\mathbf{B}_{\textrm{T}},\mathbf{B}_{\textrm{R}}\right)\lambda_{i}\left(\mathbf{B}_{\textrm{T}},\mathbf{B}_{\textrm{R}}\right)}{\sigma^{2}}\right)\label{eq:objective capacity CDB WF}\\
\text{s.t.}\ \ \  & \left[\mathbf{B}_{\textrm{T}}\right]_{:,n_{\textrm{T}}}\in\mathcal{B},\:\forall n_{\textrm{T}},\label{eq:BT CDB WF}\\
 & \left[\mathbf{B}_{\textrm{R}}\right]_{:,n_{\textrm{R}}}\in\mathcal{B},\:\forall n_{\textrm{R}}.\label{eq:BR CDB WF}
\end{align}
Both the problem \eqref{eq:objective capacity CDB UPA}-\eqref{eq:BR CDB UPA}
and problem \eqref{eq:objective capacity CDB WF}-\eqref{eq:BR CDB WF}
can be solved by successively searching the codebook for each transmit
and receive pixel antenna.

\section{Performance Evaluation}

We evaluate the performance of SISO pixel antenna system and MIMO
pixel antenna system with the proposed antenna coding design in this
section.

\subsection{Simulation Setup and Pixel Antenna Design}

We consider a propagation environment with rich scattering, which
has a 2-D uniform power angular spectrum\footnote{Herein we consider 2-D uniform power angular spectrum for illustration.
It should be noted that other power angular spectrum also applies.}, i.e. uniform over the azimuth angle on the XOY plane, with equally
likely polarization. We set the angular resolution as $\Delta\phi=5\lyxmathsym{\textdegree}$
and thus we have $K=72$ sampled angles. For the SISO pixel antenna
system, we assume that the transmit antenna is fixed with an ideal
isotropic radiation pattern and the receive antenna uses a pixel antenna.
For the MIMO pixel antenna system, both transmit and receive antennas
use pixel antennas.

\begin{figure}[t]
\begin{centering}
\includegraphics[width=8.9cm]{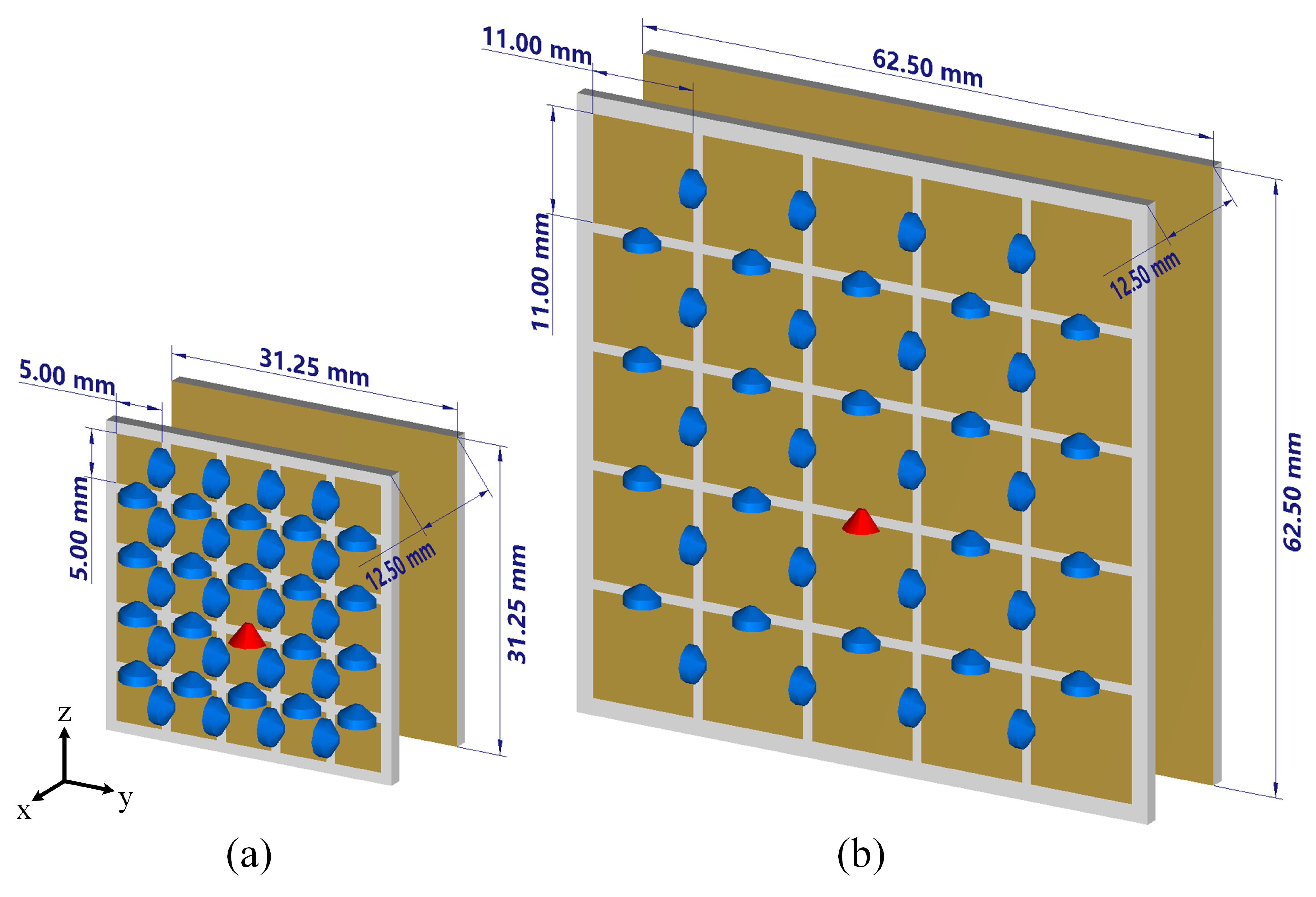}
\par\end{centering}
\caption{\label{fig:pixel antenna}Pixel antennas with different physical apertures
(a) $0.25\lambda\times0.25\lambda$ and (b) $0.5\lambda\times0.5\lambda$.
Red arrows represent the antenna ports and blue arrows represent the
pixel ports.}
\end{figure}

We consider two examples of pixel antennas operating at 2.4 GHz where
the wavelength is $\lambda=125$ mm. The two pixel antennas are designed
based on discretizing the radiating patch surface of conventional
microstrip patch antennas\footnote{Herein we consider the pixelated microstrip patch antenna for illustration.
However, it should be noted that pixel antenna is general and can
be designed based on discretizing the continuous radiating surface
of different types of antennas such as monopole antenna, not restricted
to microstrip patch antenna.} into a grid of pixels, but having different physical apertures, $0.25\lambda\times0.25\lambda$
and $0.5\lambda\times0.5\lambda$. The geometry and dimension for
the two pixel antennas are illustrated in Fig. \ref{fig:pixel antenna}
and summarized in Table \ref{tab:Geometry-and-Dimensions}. For both
pixel antennas, there are in total 40 ports, including 1 antenna port
and $Q=39$ pixel ports, and 5$\times$5 pixels with pixel size being
$5\textrm{mm}\times5\textrm{mm}$ and $11\textrm{mm}\times11\textrm{mm}$,
respectively. The pixelized patch and ground plane are printed on
the Rogers 4003C substrate, where the thickness is 1.524 mm, the permittivity
is 3.55, and the loss tangent is 0.0027. We use CST studio suite,
a full-wave EM solver, to simulate the $(Q+1)$-port circuit network
of pixel antennas to achieve the impedance matrix $\mathbf{Z}\in\mathbb{C}^{(Q+1)\times(Q+1)}$
as well as the open-circuit radiation pattern matrix $\mathbf{E}_{\textrm{oc}}\in\mathbb{C}^{2K\times(Q+1)}$.

\begin{table}[t]
\caption{\label{tab:Geometry-and-Dimensions}Geometry and Dimensions for Pixel
Antennas}

\centering{}%
\begin{tabular}{|c|c|c|}
\hline 
Physical aperture & $0.25\lambda\times0.25\lambda$ & $0.5\lambda\times0.5\lambda$\tabularnewline
\hline 
Size of ground & $\textrm{31.25\textrm{mm}\ensuremath{\times}31.25\textrm{mm}}$ & $\textrm{62.5mm\ensuremath{\times}62.5\textrm{mm}}$\tabularnewline
\hline 
Size of patch & $\textrm{29\textrm{mm}\ensuremath{\times}29\textrm{mm}}$ & $\textrm{59\textrm{mm}\ensuremath{\times}59\textrm{mm}}$\tabularnewline
\hline 
Number of pixels & 5$\times$5 & 5$\times$5\tabularnewline
\hline 
Size of pixel & $\textrm{5\textrm{mm}\ensuremath{\times}5\textrm{mm}}$ & $\textrm{11\textrm{mm}\ensuremath{\times}11\textrm{mm}}$\tabularnewline
\hline 
Gap between pixels & 1mm & 1mm\tabularnewline
\hline 
Patch and ground spacing & 12.5mm & 12.5mm\tabularnewline
\hline 
\end{tabular}
\end{table}

\begin{figure}[t]
\begin{centering}
\includegraphics[width=8.5cm]{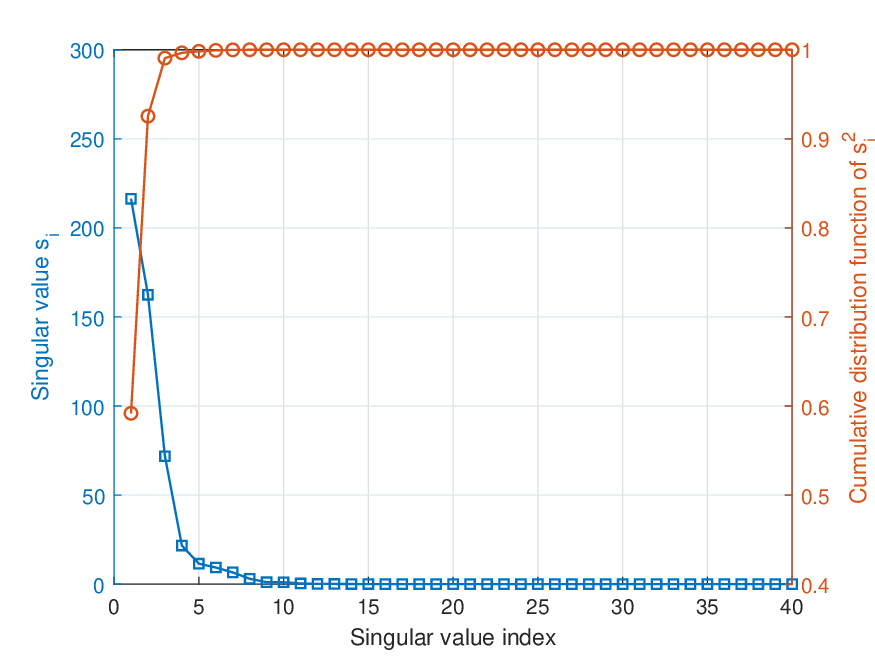}
\par\end{centering}
\begin{centering}
(a)
\par\end{centering}
\begin{centering}
\includegraphics[width=8.5cm]{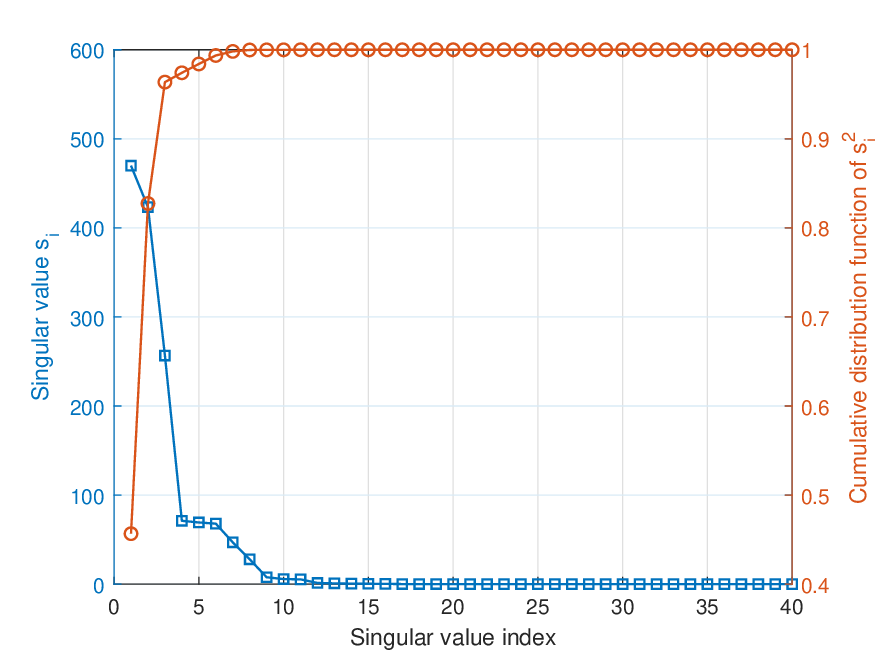}
\par\end{centering}
\begin{centering}
(b)
\par\end{centering}
\caption{\label{fig:CDF}Singular values $s_{i}$ $\forall i$ of open-circuit
radiation pattern matrix $\mathbf{E}_{\textrm{oc}}$ and cumulative
distribution function of $s_{i}^{2}$ for pixel antennas with different
physical apertures (a) $0.25\lambda\times0.25\lambda$ and (b) $0.5\lambda\times0.5\lambda$.}
\end{figure}

The singular values $s_{i}$ $\forall i$ (sorted in descending order)
of the open-circuit radiation pattern matrix $\mathbf{E}_{\textrm{oc}}$
for pixel antennas with different physical apertures are shown in
Fig. \ref{fig:CDF}. From Fig. \ref{fig:CDF}, we can observe that
the majority of the singular values are almost zero. For example,
more than half of the singular values are smaller than 0.1. Recall
that the singular value $s_{i}$ reflects the ability to radiate for
the $i$th orthogonal radiation pattern, i.e. $s_{i}^{2}$ characterizes
how much the power of the $i$th orthogonal radiation pattern is in
the total radiated power. Thus, a small value of $s_{i}^{2}$ implies
that the $i$th orthogonal radiation pattern makes negligible contribution
to the total radiation pattern of the pixel antennas. As a result,
to identify the EADoF of the pixel antennas $R=\textrm{rank}\left(\mathbf{E}_{\textrm{oc}}\right)$,
i.e. how many orthogonal radiation patterns the pixel antennas can
generate, we consider the cumulative distribution function of $s_{i}^{2}$,
defined as 
\begin{equation}
F_{i}=\frac{\sum_{j=1}^{i}s_{j}^{2}}{\sum_{j=1}^{Q+1}s_{j}^{2}},
\end{equation}
which characterizes how much power the first $i$ orthogonal radiation
patterns contribute to the total radiated power. The cumulative distribution
function of $s_{i}^{2}$ for pixel antennas with different physical
apertures are also shown in Fig. \ref{fig:CDF}. We can see that the
cumulative distribution function increases with the singular value
index and approaches unity after a certain index. Based on this, we
select the EADoF of pixel antennas as
\begin{equation}
R=\underset{F_{i}\geq T}{\textrm{argmin}}\:i,
\end{equation}
which implies that the contribution of $R$ orthogonal radiation patterns
to the total radiated power is more than the threshold $T$. In this
work, we set the threshold $T$ as 0.998, so that the EADoF of the
pixel antennas with the physical apertures $0.25\lambda\times0.25\lambda$
and $0.5\lambda\times0.5\lambda$ are 5 and 7, respectively. This
shows that the EADoF of pixel antennas increases with the physical
aperture. Moreover, recall that the upper bound for the average channel
gain of SISO pixel antenna system is given by the EADoF of pixel antennas.
The average channel gain of SISO pixel antenna system in comparison
with its upper bound will be demonstrated in the following subsection.

\subsection{SISO Pixel Antenna System Performance}

\begin{figure}[t]
\begin{centering}
\includegraphics[width=8.5cm]{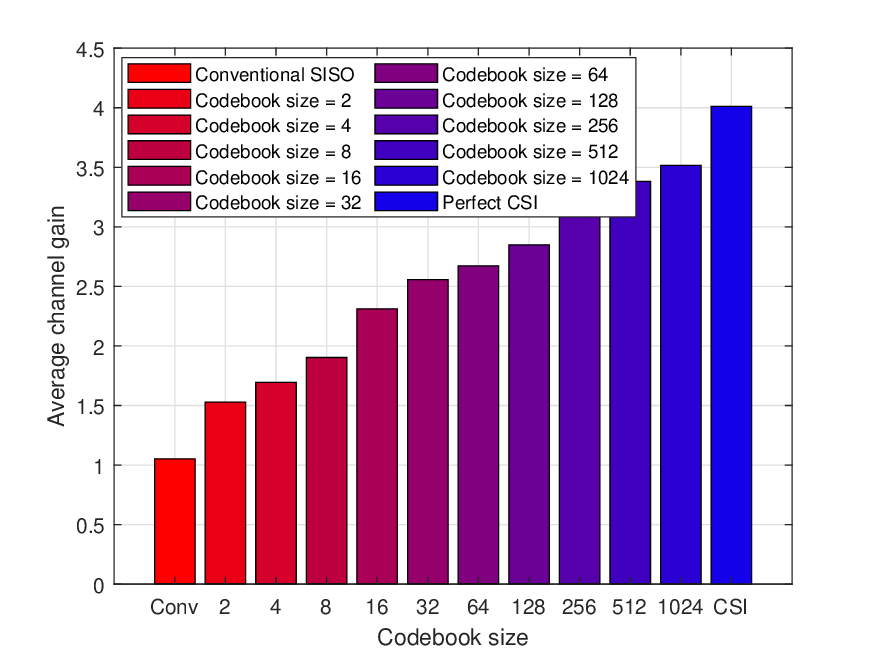}
\par\end{centering}
\begin{centering}
(a)
\par\end{centering}
\begin{centering}
\includegraphics[width=8.5cm]{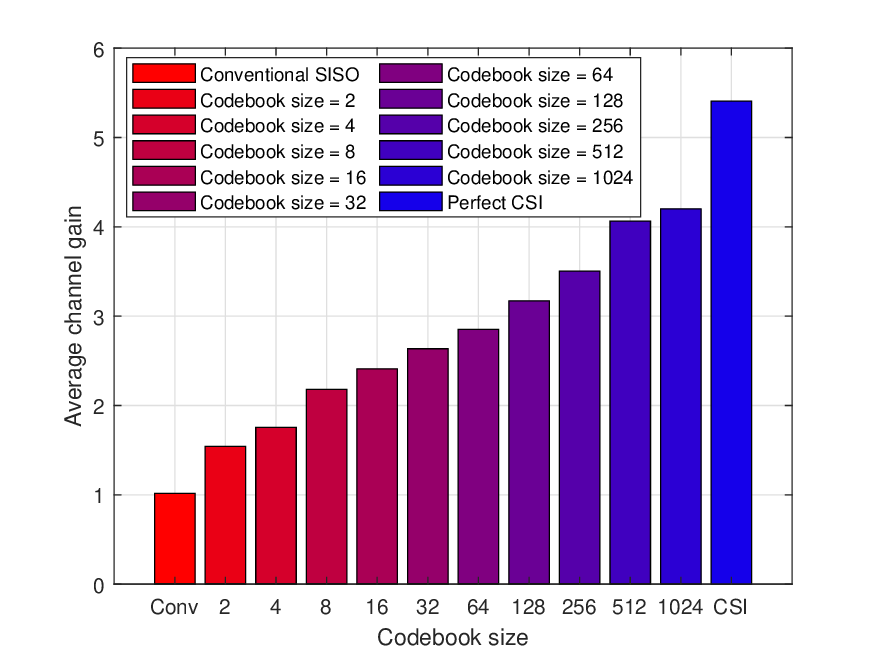}
\par\end{centering}
\begin{centering}
(b)
\par\end{centering}
\caption{\label{fig:SISO channel gain pixel0.25}Average channel gain of the
SISO pixel antenna system with physical apertures (a) $0.25\lambda\times0.25\lambda$
and (b) $0.5\lambda\times0.5\lambda$, optimized with perfect CSI
and with codebook.}
\end{figure}

We first evaluate the performance of SISO pixel antenna system. Utilizing
Monte Carlo method, we generate multiple realizations for the virtual
channel matrix $\mathbf{H}_{\textrm{V}}$ and for each realization,
we use the SEBO algorithm \cite{ShanpuShen2017_TAP_SEBO}, with block
size of $J=10$ (which achieves a good optimization performance with
an acceptable complexity), to solve the problem \eqref{eq:objective channel gain}-\eqref{eq:bR 0 1}
so as to maximize the channel gain $\left|h\left(\mathbf{b}_{\textrm{R}}\right)\right|^{2}$
with perfect CSI. In addition, we also maximize the channel gain with
the codebook design for antenna coding introduced in Section III-C.
The number of virtual channel realizations in the training set is
$L=15000$. The antenna coders in the codebook design with size\footnote{Due to limited space, it would be messy to show the antenna coders
in the codebook with size larger than 8, so we omit them.} of 2, 4, and 8 are listed in Table \ref{tab: codebook}. The number
of realizations in Monte Carlo method is 100, which is enough to obtain
an accurate average performance. The average channel gain of the SISO
pixel antenna system optimized with perfect CSI and with the codebook
is shown in Fig. \ref{fig:SISO channel gain pixel0.25}. The two pixel
antennas with different physical apertures $0.25\lambda\times0.25\lambda$
and $0.5\lambda\times0.5\lambda$ are considered and the conventional
SISO system is also compared as the benchmark. From Fig. \ref{fig:SISO channel gain pixel0.25},
we can make three observations.

\begin{table}[t]
\caption{\label{tab: codebook}Antenna Coders in the Codebook with Size of
2, 4, 8}

\centering{}%
\begin{tabular}{|c|c|c|}
\hline 
 & {\tiny Physical aperture $0.25\lambda\times0.25\lambda$} & {\tiny Physical aperture $0.5\lambda\times0.5\lambda$}\tabularnewline
\hline 
\hline 
{\tiny Codebook} & {\tiny 000110100000011101101011000001101001101} & {\tiny 010000110010110101101011001010011001000}\tabularnewline
{\tiny size = 2} & {\tiny 101010111110110111010001001111101010101} & {\tiny 100010101010010110110011100110000001101}\tabularnewline
\hline 
 & {\tiny 101001011100010111110010000110110001000} & {\tiny 001010110001100101101100100100000101010}\tabularnewline
{\tiny Codebook} & {\tiny 110111011110010100110110111000101010110} & {\tiny 111000011111110101000001000110100101111}\tabularnewline
{\tiny size = 4} & {\tiny 110100000010001000110111110000011011110} & {\tiny 100101111010010100101000011010001101110}\tabularnewline
 & {\tiny 010110001001000101111011010111111101110} & {\tiny 100101000010001001000010000101111100101}\tabularnewline
\hline 
 & {\tiny 000000111110010010110111101001011101101} & {\tiny 111111010000001000101111110100110110000}\tabularnewline
 & {\tiny 000111000011011101111000110111010001011} & {\tiny 000001101000100110101010101100001000011}\tabularnewline
 & {\tiny 010011110101110101100101111011100100001} & {\tiny 111111000111000100011000100000110001011}\tabularnewline
{\tiny Codebook} & {\tiny 001011100000010001110110100110110101100} & {\tiny 010000110101010111010100100001001100000}\tabularnewline
{\tiny size = 8} & {\tiny 010111010110011100011011111011111101010} & {\tiny 100100100010000000010110000101111000011}\tabularnewline
 & {\tiny 000100101101010000110110100110000101001} & {\tiny 010111110100011100001111100100010100111}\tabularnewline
 & {\tiny 001010011111100011010111101101100000000} & {\tiny 000101001111111101010101110100111100011}\tabularnewline
 & {\tiny 100001010010000011111000110010010111110} & {\tiny 110000011011011000100001010010100001001}\tabularnewline
\hline 
\end{tabular}
\end{table}

\begin{figure*}[tbh]
\begin{centering}
\includegraphics[width=7.5cm]{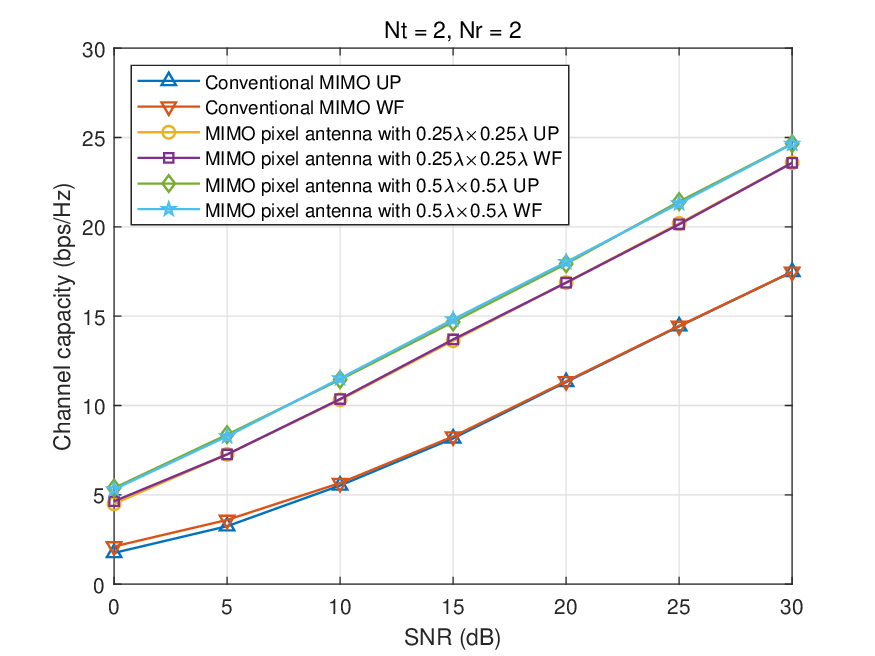}\includegraphics[width=7.5cm]{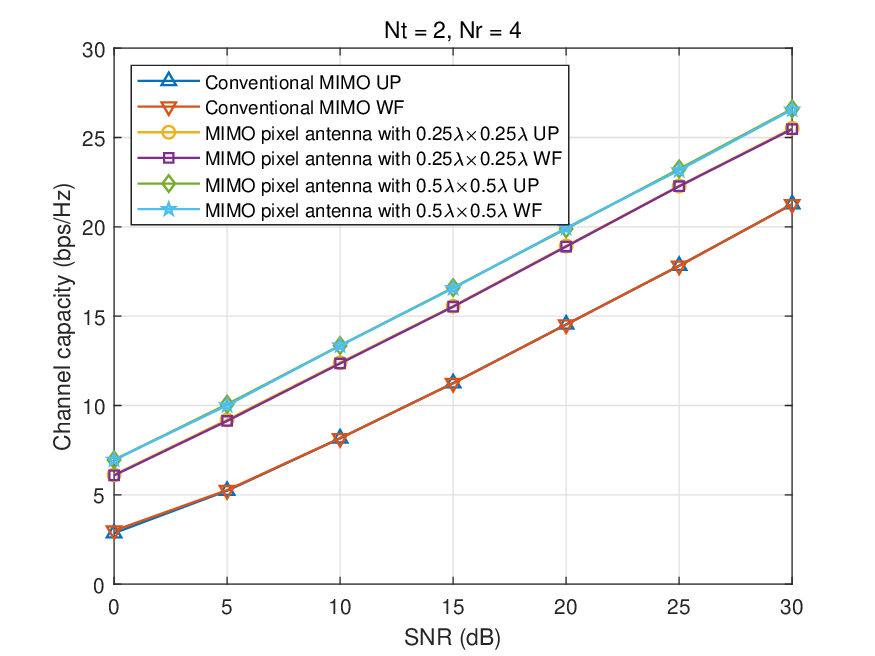}
\par\end{centering}
\begin{centering}
\includegraphics[width=7.5cm]{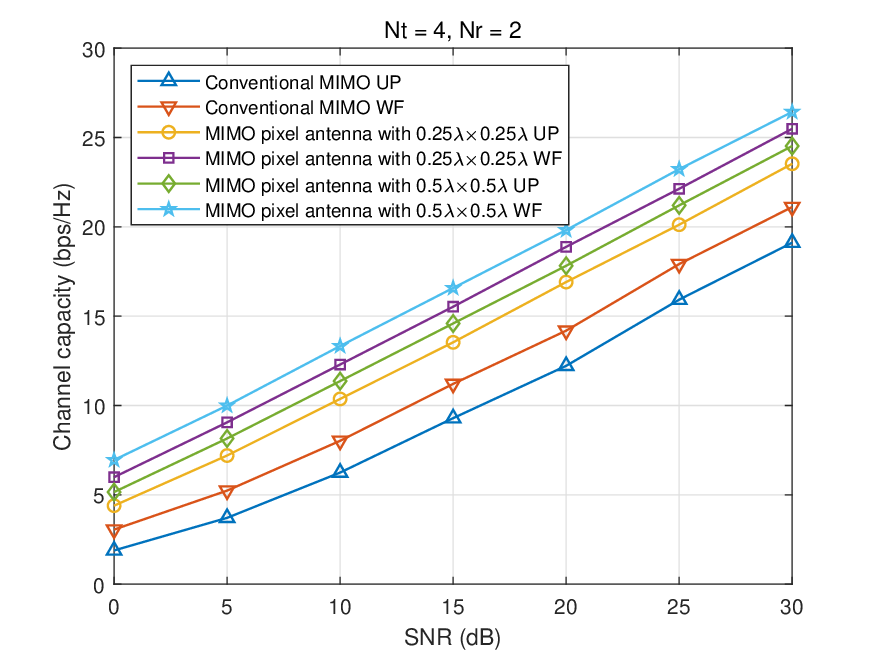}\includegraphics[width=7.5cm]{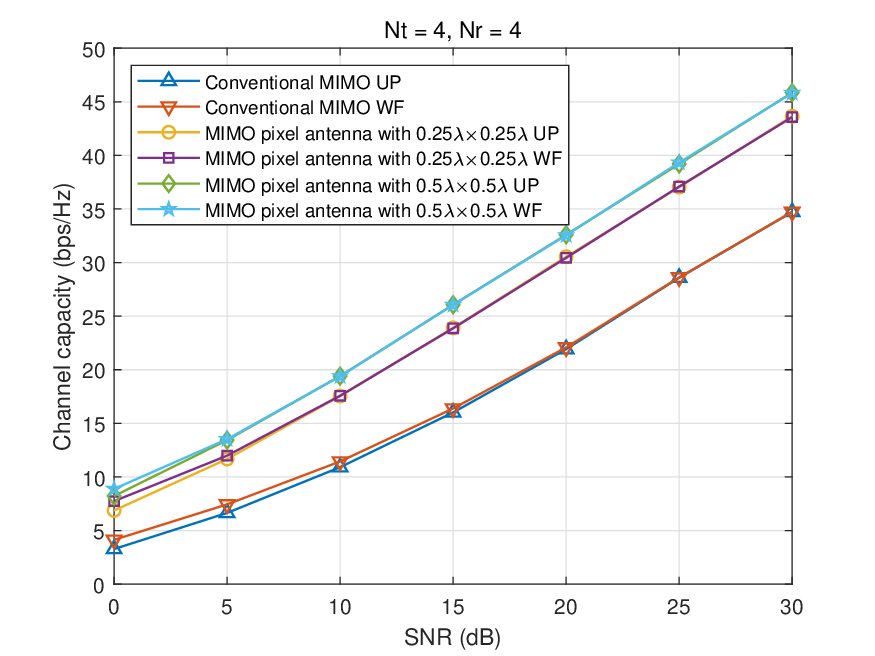}
\par\end{centering}
\caption{\label{fig:MIMO capacity}Channel capacity of the conventional MIMO
system and the MIMO pixel antenna system with the uniform power allocation
(UP) and waterfilling (WF) optimized with perfect CSI.}
\end{figure*}

\textit{Firstly}, compared with the conventional SISO system which
has a unit average channel gain from the complex Gaussian distribution
$\mathcal{CN}\left(0,1\right)$, the SISO pixel antenna system can
significantly enhance the channel gain, demonstrating the benefits
of pixel antennas. Specifically, the average channel gain with physical
apertures $0.25\lambda\times0.25\lambda$ and $0.5\lambda\times0.5\lambda$
optimized with the perfect CSI are 4 and 5.4, respectively. The channel
gain enhanced by SISO pixel antenna system can be explained by noting
that the pixel antenna can flexibly adjust its configuration so that
its radiation pattern can adapt to the virtual channel matrix to coherently
add the multiple paths in the propagation environment.

\textit{Secondly}, the average channel gain of SISO pixel antenna
system optimized with the codebook increases with the codebook size.
This is because increasing the codebook size can generate more antenna
coders, so that the pixel antenna becomes more flexible and provides
more diverse radiation patterns to enhance the channel. When the codebook
size is large enough, the average channel gain optimized with the
codebook provides performance similar to that optimized with perfect
CSI. However the computational complexity for optimization can be
significantly reduced. Particularly, even when the codebook size is
only 2, utilizing the codebook design can still enhance the average
channel gain by around 50\% for different physical apertures. Therefore,
this demonstrates the benefit of codebook design for pixel antennas.

\textit{Thirdly}, for the SISO pixel antenna system, increasing the
physical aperture is beneficial to enhance channel gain. For example,
when optimized with perfect CSI, the channel gain with physical aperture
$0.5\lambda\times0.5\lambda$ is 35\% higher than that with $0.25\lambda\times0.25\lambda$.
This is because pixel antennas with larger physical aperture can provide
more orthogonal radiation patterns, i.e. larger EADoF. Specifically,
the EADoF for the pixel antenna with physical aperture $0.25\lambda\times0.25\lambda$
is 5 while the EADoF with $0.5\lambda\times0.5\lambda$ is 7. Therefore,
pixel antennas with larger physical aperture achieve a higher upper
bound for the average channel gain \eqref{eq:upper bound}, resulting
in a higher average channel gain. Moreover, it is worth noting that
the upper bound of the average channel gain of SISO pixel antenna
system \eqref{eq:upper bound} is generally hard to achieve. For example,
for the physical aperture $0.25\lambda\times0.25\lambda$, the average
channel gain optimized with perfect CSI is 4 while the upper bound
is 5. This is because it is not guaranteed to find the antenna coder
$\mathbf{b}_{\textrm{R}}$ such that the pixel antenna current $\mathbf{i}_{\textrm{R}}\left(\mathbf{b}_{\textrm{R}}\right)$
and the $\mathbf{w}\left(\mathbf{b}_{\textrm{R}}\right)$ given by
\eqref{eq:equivalent combiner} satisfy the condition of maximum ratio
combiner \eqref{eq:MRC}.

Overall, we have shown that using pixel antennas is beneficial to
enhance the channel gain of SISO system.

\subsection{MIMO Pixel Antenna System Performance}

\begin{figure*}[tbh]
\begin{centering}
\includegraphics[width=7.5cm]{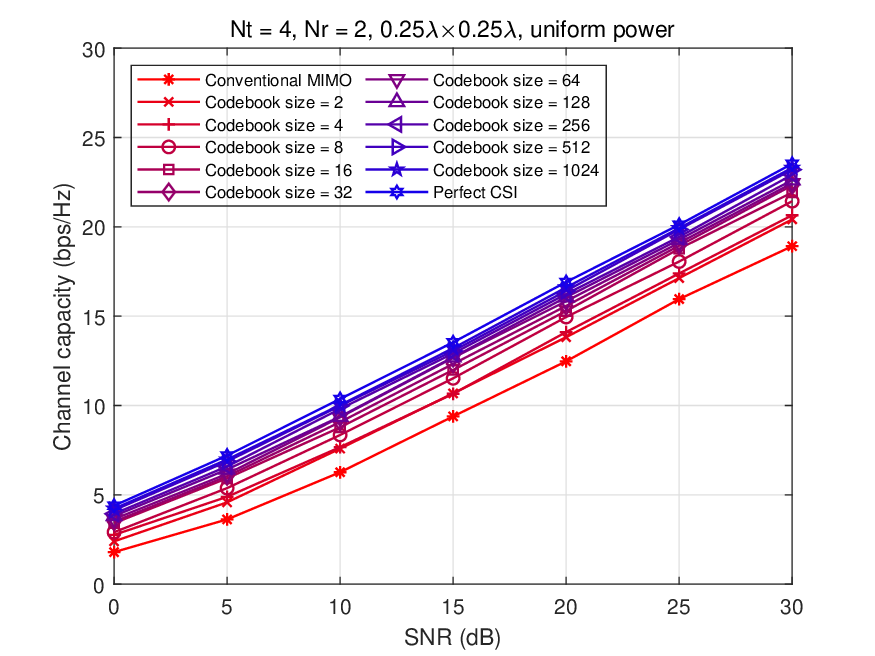}\includegraphics[width=7.5cm]{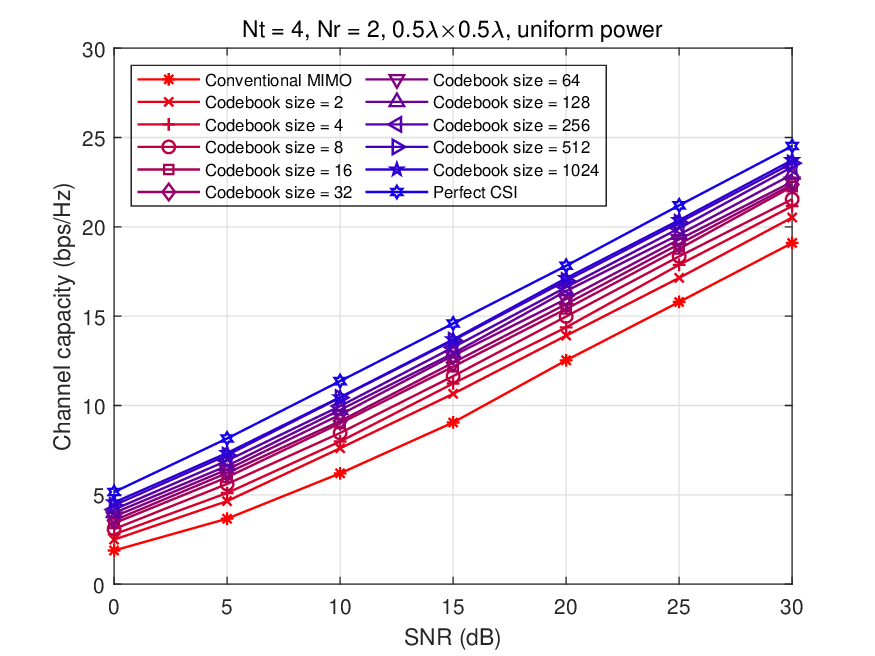}
\par\end{centering}
\begin{centering}
\includegraphics[width=7.5cm]{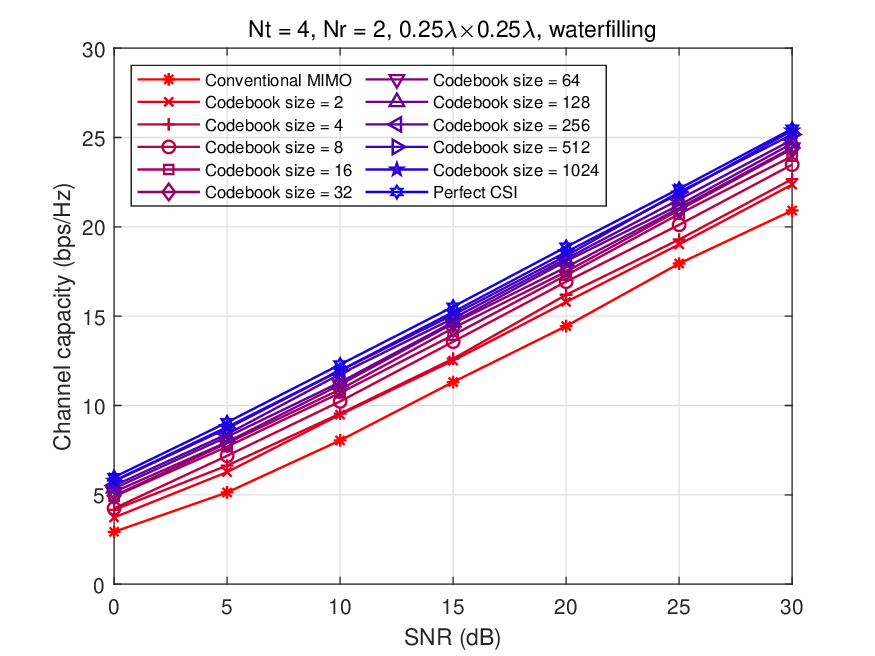}\includegraphics[width=7.5cm]{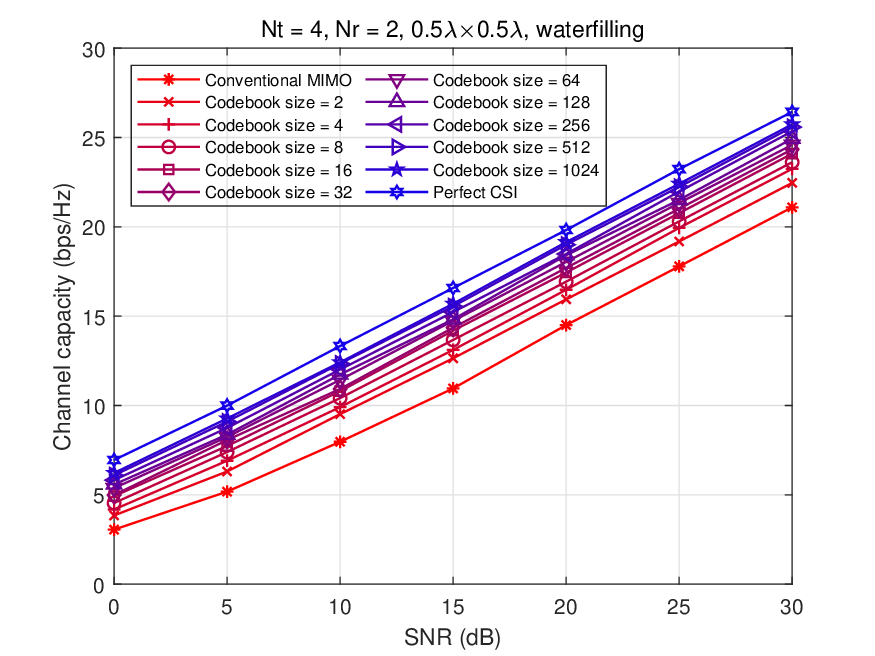}
\par\end{centering}
\caption{\label{fig:MIMO Capacity CDB}Channel capacity of the MIMO pixel antenna
system optimized with codebook.}
\end{figure*}

We next evaluate the performance of MIMO pixel antenna systems. Similarly,
we utilize the Monte Carlo method to generate multiple realizations
for the virtual channel matrix $\mathbf{H}_{\textrm{V}}$. For each
realization, we optimize the antenna coding design with uniform power
allocation by solving the problem \eqref{eq:objective capacity maximization UPA}-\eqref{eq:BR=00003D0,1 UPA}
and also jointly optimize the antenna coding and waterfilling design
by solving the problem \eqref{eq:objective capacity maximization WF}-\eqref{eq:BR=00003D0,1 WF},
for maximizing the channel capacity of MIMO pixel antenna system with
perfect CSI. The SEBO algorithm \cite{ShanpuShen2017_TAP_SEBO}, with
block size $J=10$, is utilized to solve the two problems. The channel
capacity of the MIMO pixel antenna system with uniform power allocation
(UP) and waterfilling (WF) optimized with perfect CSI is shown in
Fig. \ref{fig:MIMO capacity}. The channel capacity is averaged over
multiple channel realizations. The two pixel antennas with different
physical apertures $0.25\lambda\times0.25\lambda$ and $0.5\lambda\times0.5\lambda$
are considered and the conventional MIMO system with UP and WF are
compared. Different numbers of transmit and receive antennas $N_{\textrm{T}}$
and $N_{\textrm{R}}$ are also considered for the MIMO pixel antenna
system and conventional MIMO system. From Fig. \ref{fig:MIMO capacity},
we can make four observations.

\textit{Firstly}, in comparison with the conventional MIMO system,
the MIMO pixel antenna system can effectively enhance channel capacity,
demonstrating the benefits of pixel antennas. For example, in the
case of $N_{\textrm{T}}=2$ and $N_{\textrm{R}}=2$, the channel capacity
with UP at low SNR of 0 dB for the conventional MIMO system and the
MIMO pixel antenna system with physical apertures $0.25\lambda\times0.25\lambda$
and $0.5\lambda\times0.5\lambda$ is 1.75, 4.50, and 5.38 bps/Hz,
respectively, which implies that the capacity for MIMO pixel antenna
systems are 2.6 and 3.1 times higher than that for the conventional
MIMO system. On the other hand, in the same case but at high SNR of
30 dB, the channel capacity with UP for the conventional MIMO system
and the MIMO pixel antenna system with physical apertures $0.25\lambda\times0.25\lambda$
and $0.5\lambda\times0.5\lambda$ are 17.48, 23.58 and 24.65 bps/Hz,
respectively, which means that using pixel antenna can enhance the
capacity by 35\% and 41\%. The channel capacity enhanced by MIMO pixel
antenna system occurs because the multiple pixel antennas at the transmitter
and receivers can flexibly adjust their radiation patterns to adapt
to the virtual channel matrix $\mathbf{H}_{\textrm{V}}$ so that the
multiple paths with different AoA and AoD in the propagation environment
can be coherently added to enhance the channel $\mathbf{H}\left(\mathbf{B}_{\textrm{T}},\mathbf{B}_{\textrm{R}}\right)$
in \eqref{eq:beamspace MIMO pixel antenna} as well as the capacity.

\textit{Secondly}, increasing the physical aperture is beneficial
to enhance the channel capacity of MIMO pixel antenna system. For
example, at low SNR of 0 dB, the channel capacity with UP for MIMO
pixel antenna system with physical apertures $0.5\lambda\times0.5\lambda$
is 20\% higher than that with $0.25\lambda\times0.25\lambda$. This
is because pixel antennas with larger physical aperture provide larger
EADoF which increases the channel gain between each pair of transmit
and receive pixel antennas and therefore leads to higher channel capacity.

\textit{Thirdly}, for MIMO pixel antenna system, increasing the number
of transmit antennas and receive antennas is beneficial to improve
the channel capacity. For example, with the same physical aperture,
the channel capacity at high SNR of 30 dB for the $4\times4$ MIMO
pixel antenna system is nearly twice that for the $2\times2$ MIMO
pixel antenna system, which is consistent with the conventional MIMO
system. That is to say, utilizing pixel antennas to construct MIMO
systems will not affect the multiplexing gain of MIMO system. Moreover,
it will bring extra channel gain on top of the multiplexing gain.

\textit{Fourthly}, the joint antenna coding and waterfilling design
achieves higher channel capacity than antenna coding design with uniform
power allocation, especially when there are more transmit antennas
than receive antennas. For example, for the $4\times2$ MIMO pixel
antenna system with physical aperture $0.25\lambda\times0.25\lambda$,
the channel capacity with UP and WF is 4.39 and 5.99 bps/Hz at low
SNR of 0 dB, respectively. This implies that joint antenna coding
and waterfilling design can increase the capacity by 36\%. Thus, this
shows that pixel antennas can synergize with waterfilling design to
enhance channel capacity in MIMO systems.

We also use the codebook to optimize the antenna coding design with
uniform power allocation and waterfilling to maximize the channel
capacity of MIMO pixel antenna system. Specifically, we solve the
problem \eqref{eq:objective capacity CDB UPA}-\eqref{eq:BR CDB UPA}
and problem \eqref{eq:objective capacity CDB WF}-\eqref{eq:BR CDB WF}
by successively searching the codebook for each transmit and receive
pixel antenna. The channel capacity of the MIMO pixel antenna system
with UP and WF optimized with the codebook is shown in Fig. \ref{fig:MIMO Capacity CDB}.
The channel capacity is averaged over channel realizations and different
physical apertures are considered (we only show the $4\times2$ case
for brevity). From Fig. \ref{fig:MIMO Capacity CDB}, we can make
two observations.

\textit{Firstly}, the channel capacity of MIMO pixel antenna system
optimized with the codebook increases with the physical aperture.
Joint antenna coding and waterfilling design achieves higher channel
capacity than antenna coding design with uniform power allocation.
These are consistent with the case optimized with CSI.

\textit{Secondly}, the channel capacity of MIMO pixel antenna system
optimized with the codebook increases with the codebook size, which
is because increasing the codebook size gives the pixel antenna more
diverse radiation patterns to enhance the channel. When the codebook
size is large enough, the capacity optimized with codebook is similar
to that optimized with perfect CSI but the computational complexity
can be reduced. Even when the codebook size is only 2, using the codebook
design is still beneficial. For example, the capacity with UP for
the $4\times2$ MIMO pixel antenna system at low SNR of 0 dB is 30\%
higher than that for conventional MIMO system when codebook size is
2. Thus, this shows the benefit and effectiveness of codebook for
MIMO pixel antenna system.

Overall, we have shown that using pixel antennas is beneficial to
enhance the channel capacity of MIMO system.

\subsection{Evaluation of Multiport Circuit Network Model}

\begin{figure}[t]
\begin{centering}
\includegraphics[width=4cm]{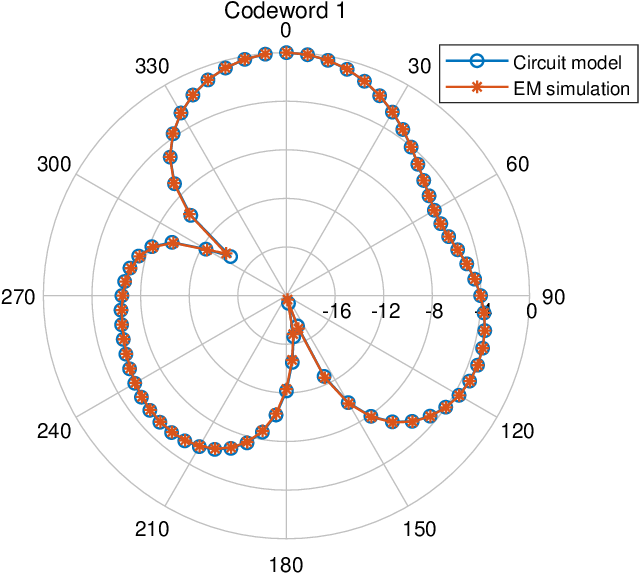}\includegraphics[width=4cm]{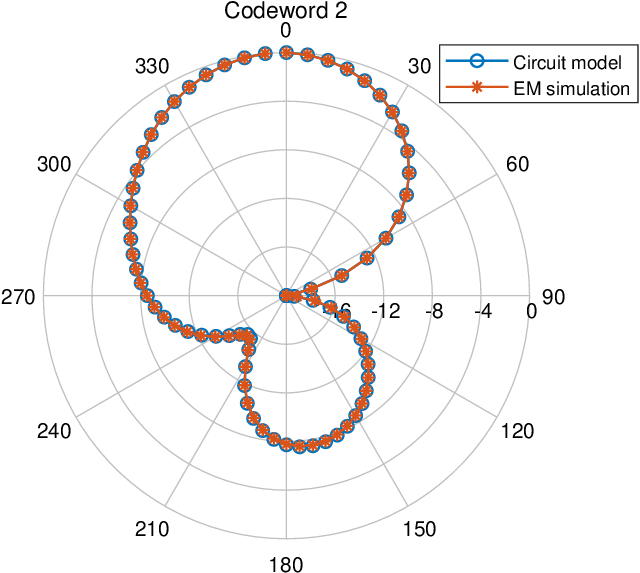}
\par\end{centering}
\begin{centering}
\includegraphics[width=4cm]{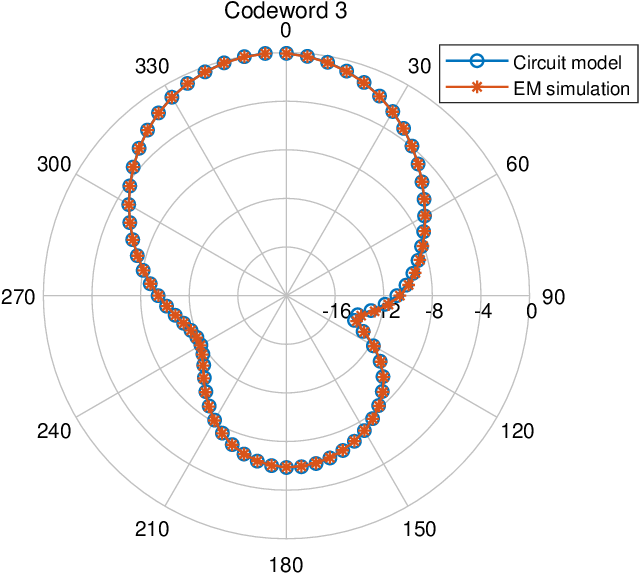}\includegraphics[width=4cm]{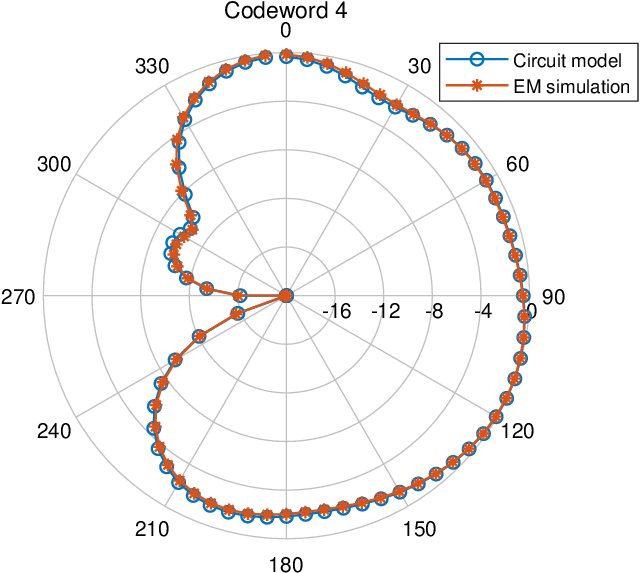}
\par\end{centering}
\caption{\label{fig:Radiation-patterns}Radiation patterns of the pixel antenna
(with physical aperture $0.5\lambda\times0.5\lambda$) configured
by antenna coders in the codebook (with codebook size of 4) using
EM simulation and multiport circuit network model. Unit: dB.}
\end{figure}

To validate the accuracy of the multiport circuit network model, we
compare it with the full-wave EM simulation using CST studio suite.
Specifically, in CST studio suite, we simulate the pixel antenna with
physical aperture $0.5\lambda\times0.5\lambda$ configured by different
antenna coders in the codebook (with codebook size of 4). The radiation
patterns of the pixel antenna, obtained by the EM simulation and the
multiport circuit network model \eqref{eq:F pattern} and \eqref{eq: current},
are shown in Fig. \ref{fig:Radiation-patterns}. We can find that
the EM simulated radiation patterns are almost same as those computed
by the multiport circuit network model for different antenna coders,
showing the accuracy of the proposed model. In addition, the radiation
patterns for different antenna coders are diverse, covering different
angle ranges, which is consistent with the fact that the codebook
is designed to maximize the ergodic channel gain in a multipath propagation
environment with 2-D uniform power angular spectrum.

More importantly, under the same accuracy, the multiport circuit network
model is significantly more computationally efficient. We measure
the computational time to calculate the radiation pattern using the
EM simulation (CST studio suite) and the proposed multiport circuit
network model, respectively, in a laptop (with AMD Ryzen 9 7940HX
CPU and 16 GB memory). The average computational time for the EM simulation
and the multiport circuit network model, denoted as $T_{\textrm{EM}}$
and $T_{\textrm{circuit}}$, is measured around 56.59s and 0.000535s,
respectively, which shows that the proposed model is around $10^{5}$
time faster than the EM simulation.

Note that we only need one EM simulation to obtain $\mathbf{Z}$ and
$\mathbf{E}_{\textrm{oc}}$ for the multiport circuit network model.
The computational time to simulate $\mathbf{Z}$ and $\mathbf{E}_{\textrm{oc}}$
for a realistic pixel antenna is acceptable as the number of pixels
is generally not too large with the size of pixel around 0.1-0.2$\lambda$,
which is small enough to achieve high reconfigurability\footnote{Further reducing the size of pixel can bring marginal gain but at
the expense of high circuit complexity to control massive switches.
Thus, it is unnecessary to have a very large number of pixels with
very small pixels.}. In this work, the computational time of the EM simulation to obtain
$\mathbf{Z}$ and $\mathbf{E}_{\textrm{oc}}$, denoted as $T_{0}$,
is measured around 12 minutes\footnote{When using CST studio suite to simulate the multiport circuit network
of pixel antenna, the $(Q+1)\times(Q+1)$ impedance matrix $\mathbf{Z}$
and the open-circuit radiation patterns for the $Q+1$ ports $\mathbf{E}_{\textrm{oc}}$
can be obtained by performing only one EM simulation. Such one EM
simulation contains ($Q+1$) sub-simulations, so that each port will
be sequentially excited and simulated in each sub-simulation. Thus,
the total computational time to obtain $\mathbf{Z}$ and $\mathbf{E}_{\textrm{oc}}$
(including sequentially exciting each port in each sub-simulation)
is 12 minutes.} by the same laptop above. In spite of such time, the EM simulation
can be done offline and once $\mathbf{Z}$ and $\mathbf{E}_{\textrm{oc}}$
are known the radiation pattern can be calculated accurately and efficiently
for real-time antenna coding optimization.

Last but not least, we completely compare the computational time for
the whole process to optimize the antenna coder using the proposed
multiport circuit network model and the EM simulation in the following
two cases.

\textit{First}, when using SEBO to optimize the antenna coder, i.e.
solving the problem \eqref{eq:objective channel gain}-\eqref{eq:bR 0 1},
the total number of calculations of radiation pattern is $N_{\textrm{e}}2^{J}$,
so the total computational time for the proposed multiport circuit
network model is
\begin{equation}
T_{\textrm{total,circuit}}=T_{0}+N_{\textrm{e}}2^{J}T_{\textrm{circuit}},
\end{equation}
while the total computational time using EM simulation is
\begin{equation}
T_{\textrm{total,EM}}=N_{\textrm{e}}2^{J}T_{\textrm{EM}}.
\end{equation}
If considering $N_{\textrm{e}}=4$ (the least number to optimize all
the $Q$ switches), we have $T_{\textrm{total,circuit}}=\textrm{722}\,\textrm{s}$
and $T_{\textrm{total,EM}}=\textrm{231792}\,\textrm{s}.$

\textit{Second}, when using the codebook to optimize the antenna coder,
we need to design the codebook, which needs $N_{\textrm{c}}=N_{\textrm{o}}\left(\sum_{m=1}^{M}N_{\textrm{e},m}2^{J}\right)$
calculations of radiation pattern, where $N_{\textrm{o}}$ is the
number of iterations of antenna coder optimization in GLA and $N_{\textrm{e},m}$
is the number of iterations for using SEBO to optimize the $m$th
antenna coder in codebook with size of $M$. Once the codebook is
ready, we search the codebook by $M$ calculations of radiation pattern.
Totally, we need $N_{\textrm{c}}+M$ calculations of radiation pattern
to use codebook to optimize antenna coder, so the total computational
time for the proposed multiport circuit network model is
\begin{equation}
T_{\textrm{total,circuit}}=T_{0}+\left(N_{\textrm{o}}\left(\sum_{m=1}^{M}N_{\textrm{e},m}2^{J}\right)+M\right)T_{\textrm{circuit}},
\end{equation}
while the total computational time using EM simulation is
\begin{equation}
T_{\textrm{total,EM}}=\left(N_{\textrm{o}}\left(\sum_{m=1}^{M}N_{\textrm{e},m}2^{J}\right)+M\right)T_{\textrm{EM}}.
\end{equation}
If considering $N_{\textrm{o}}=1$ (the least number to perform antenna
coding optimization in GLA), $N_{\textrm{e},m}=4$ $\forall m$, and
e.g. $M=16$, we have $T_{\textrm{total,circuit}}=\textrm{755}\,\textrm{s}$
and $T_{\textrm{total,EM}}=\textrm{3709587}\,\textrm{s}$.

From the above complete comparison, we have shown that the proposed
multiport circuit network model is significantly more computationally
efficient than the EM simulation.

\section{Discussions and Future Work}

\subsection{Continuous Antenna Coding}

It is possible to continuously adjust the load impedance $z_{\textrm{L},q}$
to implement a continuous antenna coding. To that end, we can use
variable reactive load, such as varactor, to replace the RF switch
in the pixel antenna, so that $z_{\textrm{L},q}$ can be modeled as
$z_{\textrm{L},q}=jx_{\textrm{L},q}$, where $j$ is the imaginary
unit and $x_{\textrm{L},q}$ is the continuous variable reactance.
Accordingly, the continuous antenna coding design becomes a continuous
optimization problem, which is easier to solve and can provide better
performance compared with the binary antenna coding. However, implementing
the continuous antenna coding increases the circuit complexity to
control the pixel antenna since continuous dc bias voltages are required.
Therefore, there is a tradeoff between the circuit complexity, system
performance, and difficulty for optimization.

\subsection{Antenna Coding Optimization for More Scenarios}

Antenna coding can be optimized for more scenarios in future research,
including, but are not limited to, the following:

\subsubsection{Multi-User MIMO Pixel Antenna System}

We can jointly optimize the precoding, combining, and antenna coding
at transmitter and users to enhance the signal-to-interference-plus-noise
ratio to maximize the weighted sum rate.

\subsubsection{OFDM MIMO Pixel Antenna System}

We can derive the frequency dependent model for pixel antenna and
subsequently jointly optimize the precoding at each carrier frequency
and the antenna coding of pixel antenna to maximize the capacity.

\subsubsection{Pixel Antenna System for Wireless Power Transfer}

We can jointly optimize the waveform, beamforming, and antenna coding
to maximize the output dc power.

\subsection{Pixel Antenna Element Spacing}

There is a limitation on the spacing between adjacent pixel antenna
elements, which should be neither too small nor too large, as explained
in the following.

\subsubsection{Inter-Element Electromagnetic Interference}

When the spacing is too small, mutual coupling between the pixel antennas
will be high and degrade the performance. In this work, we address
the inter-element electromagnetic interference (EMI) issue in the
MIMO pixel antenna system through separating the pixel antennas by
a suitable distance to suppress the mutual coupling. For example,
we separate two pixel antennas with physical aperture $0.25\lambda\times0.25\lambda$
and $0.5\lambda\times0.5\lambda$ by half-wavelength and one wavelength,
respectively. Moreover, we can extend the proposed multiport circuit
network model to take the inter-element EMI issue into account. For
example, we consider two closely placed pixel antennas, which can
be modeled as a ($2Q+2$)-port circuit network characterized by a
$(2Q+2)\times(2Q+2)$ impedance matrix $\mathbf{Z}$. Given the antenna
coders for pixel antennas 1 and 2, denoted as $\mathbf{b}_{1}$ and
$\mathbf{b}_{2}$, we can find the corresponding load impedance $\mathbf{Z}_{\textrm{L},1}\left(\mathbf{b}_{1}\right)$
and $\mathbf{Z}_{\textrm{L},2}\left(\mathbf{b}_{2}\right)$. Thus,
from $\mathbf{Z}$, $\mathbf{Z}_{\textrm{L},1}\left(\mathbf{b}_{1}\right)$,
and $\mathbf{Z}_{\textrm{L},2}\left(\mathbf{b}_{2}\right)$, we can
find the currents $\mathbf{i}_{1}\left(\mathbf{b}_{1},\mathbf{b}_{2}\right)$
and $\mathbf{i}_{2}\left(\mathbf{b}_{1},\mathbf{b}_{2}\right)$ and
radiation patterns $\mathbf{e}_{1}\left(\mathbf{b}_{1},\mathbf{b}_{2}\right)$
and $\mathbf{e}_{2}\left(\mathbf{b}_{1},\mathbf{b}_{2}\right)$ for
pixel antennas 1 and 2. Using the beamspace channel model, we can
link the radiation patterns to the channel, so that we can optimize
$\mathbf{b}_{1}$ and $\mathbf{b}_{2}$ to enhance the performance.
By this way, the inter-element EMI issue can be fully characterized
and considered. However, the extended model implies that the radiation
pattern of individual pixel antenna is jointly coded by all the antenna
coders due to the EMI issue, making the model complicate and intractable
and increasing the complexity for antenna coding optimization.

\subsubsection{Spherical Wave}

When the spacing is too large, it makes the MIMO pixel antenna system
very large, so the assumption of plane wave that we use becomes inaccurate.
In this work, we assume that pixel antenna element spacing is not
too large, as mentioned above, so that the plane wave assumption is
valid. Otherwise, the effect of spherical wave should be incorporated
in beamspace channel model for MIMO pixel antenna systems.

\subsection{Pixel Antenna System Prototyping}

Various pixel antennas have been designed, prototyped, and experimented.
For example, in \cite{7950976} and \cite{9769906}, the experiments
for the pixel antenna prototypes have verified that the pixel antennas
can flexibly reconfigure the radiation pattern toward different directions.
Considering a line-of-sight (LOS) channel scenario, such pixel antennas
can be directly used to generate an adaptive beam toward the receiver
to enhance the system performance such as channel gain. Thus, the
experiments provided in \cite{7950976} and \cite{9769906} have preliminarily
verified the advantages of pixel antenna system in a simple LOS scenario.
However, the previous work \cite{7950976} and \cite{9769906} were
investigated at the level of antenna hardware design and limited to
LOS scenario. To overcome this limitation, in this work we propose
antenna coding which considers the impact on wireless communication
with multipath fading at the system level to enhance the performance.
Thus, in future research, the proposed pixel antenna system can be
prototyped similarly to \cite{7950976} and \cite{9769906}, but needs
more efforts in the signal processing to control the pixel antenna
with the proposed antenna coding technique.

\section{Conclusions}

In this paper, we propose a novel technique denoted antenna coding
empowered by pixel antennas. Pixel antennas are a flexible reconfigurable
antenna technology based on discretizing a continuous radiation surface
into small elements termed pixels. The characteristics of the pixel
antennas, such as radiation patterns can be controlled by adjusting
the connections between adjacent pixels through switches, which can
be leveraged to enhance wireless communication systems.

We first derive a physical and EM based communication model for pixel
antennas using microwave multiport network theory and beamspace channel
representation. With the communication model, we consider the SISO
pixel antenna system and optimize antenna coding using the SEBO algorithm
to maximize channel gain. To lower computational complexity, we propose
a codebook design for antenna coding based on the generalized Lloyd
algorithm. We also analyze the average channel gain of SISO pixel
antenna systems and derive that the upper bound of average channel
gain is given by the EADoF of the pixel antennas. In addition, we
jointly optimize the antenna coding and transmit signal covariance
matrix to maximize the channel capacity in MIMO pixel antenna systems.

We evaluate the performance of SISO and MIMO pixel antenna systems
compared to conventional SISO and MIMO systems. It is shown that using
pixel antennas can enhance the average channel gain by up to 5.4 times
and the channel capacity by up to 3.1 times, demonstrating the significant
potential of pixel antennas as a new dimension to the design and optimization
of wireless communication systems.

\bibliographystyle{IEEEtran}
\bibliography{IEEEabrv,IEEEexample,FAS,Wireless,ShanpuShen,Pixel_antenna,Reference}

\end{document}